\documentclass[usenatbib]{mnras}
\usepackage{graphicx}
\usepackage{float}
\usepackage{amssymb}
\usepackage{color}
\usepackage{multirow}
\usepackage[T1]{fontenc}
 
\newcommand{\sarc}{$^{\prime\prime}\!\!$}

\newcommand{\myreferences}{ms}

\def\aj{AJ}                  
\def\apj{ApJ}                
\def\apjl{ApJ}               
\def\apjs{ApJS}              
\def\ao{Appl.Optics}         
\def\aap{A\&A}               
\def\aapr{A\&A~Rev.}         
\def\aaps{A\&AS}             
\def\mnras{MNRAS}            
\def\pasj{PASJ}              
\def\nat{Nature}             

\catcode`~=11 
\newcommand{\urltilde}{\kern -.15em\lower .7ex\hbox{~}\kern .04em}
\catcode`~=13 

\title[4C 43.15 at 55$\,$MHz]{LOFAR  VLBI Studies at 55$\,$MHz of 4C 43.15, a z=2.4 Radio Galaxy}
\author[Leah K. Morabito]{\parbox{\textwidth}{Leah K. Morabito$^{1}$\thanks{E-mail: morabito@strw.leidenuniv.nl}, Adam T. Deller$^{2}$, Huub R\"{o}ttgering$^{1}$, George Miley$^{1}$, Eskil Varenius$^{3}$, Timothy W. Shimwell$^{1}$,  Javier Mold\'{o}n$^{4}$,  Neal Jackson$^{3}$,  Raffaella Morganti$^{2,5}$,  Reinout J. van Weeren$^{6}$, J.~B.~R. Oonk$^{1,2}$ \\}\\
$^{1}$Leiden Observatory, P.O. Box 9513, 2300 RA, Leiden, The Netherlands \\
$^{2}$Netherlands Institute for Radio Astronomy (ASTRON), Postbus 2, 7990 AA Dwingeloo, The Netherlands \\
$^{3}$Department of Earth and Space Sciences, Chalmers University of Technology, Onsala Space Observatory, 439 92 Onsala, Sweden \\
$^{4}$Jodrell Bank Centre for Astrophysics, School of Physics and Astronomy, University of Manchester, Oxford Road, Manchester M13 9PL, UK\\
$^{5}$Kapteyn Astronomical Institute, University of Groningen, P.O. Box 800, 9700 AV Groningen, The Netherlands \\
$^{6}$Harvard-Smithsonian Center for Astrophysics, 60 Garden Street, Cambridge, MA 02138, USA \\}

\definecolor{Mygrey}{gray}{0.75}

\newcommand{\mjy}[1]{$\,$mJy$\,$bm$^{-1}$}

\begin{document}

\date{}

\pagerange{\pageref{firstpage}--\pageref{lastpage}} \pubyear{2016}

\maketitle

\label{firstpage}

\begin{abstract}
The correlation between radio spectral index and redshift has been exploited to discover high redshift radio galaxies, but its underlying cause is unclear. It is crucial to characterise the particle acceleration and loss mechanisms in high redshift radio galaxies to understand why their radio spectral indices are steeper than their local counterparts. Low frequency information on scales of $\sim$ 1 arcsec are necessary to determine the internal spectral index variation. 
In this paper we present the first spatially resolved studies at frequencies below 100$\,$MHz of the $z=2.4$ radio galaxy 4C 43.15 which was selected based on its ultra-steep spectral index ($\alpha<-1$; $S_{\nu}\sim\nu^{\alpha}$) between 365$\,$MHz and 1.4$\,$GHz. Using the International Low Frequency Array (LOFAR) Low Band Antenna we achieve sub-arcsecond imaging resolution at 55$\,$MHz with VLBI techniques. 
Our study reveals low-frequency radio emission extended along the jet axis, which connects the two lobes. 
The integrated spectral index for frequencies $<500\,$MHz is -0.83. The lobes have integrated spectral indices of -1.31$\pm0.03$ and -1.75$\pm0.01$ for frequencies $\geq1.4\,$GHz, implying a break frequency between $500\,$MHz and $1.4\,$GHz. These spectral properties are similar to those of local radio galaxies. We conclude that the initially measured ultra-steep spectral index is due to a combination of the steepening spectrum at high frequencies with a break at intermediate frequencies. 
\end{abstract}


\section{Introduction}
High redshift radio galaxies (HzRGs) are rare, spectacular objects with extended radio jets whose length exceeds scales of a few kiloparsecs. The radio jets are edge-brightened, Fanaroff-Riley class II \citep[FR$\,$II;][]{fr74} sources. Found in overdensities of galaxies indicative of protocluster environments \citep[e.g.,][]{pent00}, HzRGs are among the most massive galaxies in the distant universe and are likely to evolve into modern-day dominant cluster galaxies \citep{mdb08,blr97}. They are therefore important probes for studying the formation and evolution of massive galaxies and clusters at $z\geq2$. 

One of the most intriguing characteristics of the relativistic plasma in HzRGs is the correlation between the steepness of the radio spectra and the redshift of the associated host galaxy \citep{tielens79,bm79}.
Radio sources with steeper spectral indices are generally associated with galaxies at higher redshift, and samples of radio sources with ultra-steep spectra ($\alpha \lesssim -1$ where the flux density $S$ is $S\propto\nu^{\alpha}$) were effectively exploited to discover HzRGs \citep[e.g.,][]{rott94,chambers90,cmb87}.  

 The underlying physical cause of this relation is still not understood. Three causes  have been proposed: observational biases, environmental influences, and internal particle acceleration mechanisms that produce intrinsically steeper spectra. 
 
Several observational biases can impact the measured relation. \cite{klamer06} explored the radio \textquotedblleft $k$-correction\textquotedblright\ using a sample of 28 spectroscopically confirmed HzRGs. The authors compared the relation between spectral index and redshift as measured from the observed and rest frame spectra, and found that the relation remained unchanged. Another bias could come from the fact that jet power and spectral index are correlated. This manifests in an observed luminosity$-$redshift correlation: brighter sources (which tend to be at higher redshifts) are more likely to have higher jet power, and therefore steeper spectral indices. For flux density limited surveys this leads to a correlation between power and redshift, and surveys with higher flux density limits have a tighter power$-$redshift correlation \citep{blundell99}. 

Environmental effects could also impact the relation. The temperature of the circumgalactic medium is expected to be higher at higher redshifts. It is also known that the linear sizes of radio sources decrease with redshift \citep[e.g.,][]{miley68,neeser95} which is interpreted as lower expansion speeds due to higher surrounding gas densities at higher redshifts. \citet{ak98} point out that the expanding radio lobes therefore have to work against higher density and temperature. This would slow down the propagation of the jet into the medium, increasing the Fermi acceleration and thus steepening the spectral index. The power$-$redshift correlation in this case would be caused by a change in environment with redshift. 

The final option is that the steeper spectrum is indicative of particle acceleration mechanisms different from those in local radio galaxies. One global difference between low and high redshift sources is that the CMB temperature is higher, and could provide more inverse Compton losses at high frequencies from scattering with CMB photons. Internally to a radio galaxy, spectral indices are seen to evolve along the radio jet axis, with hot spots dominant at high frequencies, and diffuse lobe emission is dominant at low frequencies \citep[e.g. Cygnus A;][]{carilli91}.
 Recently \citet{mckean16} observed a turnover in the spectra of the hot spots detected with LOFAR around 100$\,$MHz. The authors were able to rule out a cut off in the low-energy electron distribution, and found that both free-free absorption or synchrotron self-absorption models provided adequate fits to the data, albeit with unlikely model parameters. To determine the particle acceleration mechanisms it is crucial to make observations at 100$\,$MHz and below with sufficient resolution to determine the internal variation of the low-frequency spectra.  This can then be compared to archival observations with similar or higher resolution at frequencies above 1$\,$GHz, where the internal structure of HzRGs have been well studied \citep[e.g.,][]{carilli97,pentVLA00}. All current low frequency information that does exist comes from studies in which HzRGs are unresolved. 

Typical angular sizes of HzRGs with $z\gtrsim2$ are about $10\,\textrm{arcsec}$ \citep{wm74}, driving the need for resolutions of about an arcsecond to determine the distribution of spectral indices among spatially resolved components of HzRGs.
The unique capabilities of the Low Frequency Array \citep[LOFAR;][]{vh13} are ideally suited for revealing these distributions at low frequencies. Covering the frequency bands of $10$--$80\,\textrm{MHz}$ (Low Band Antenna; LBA) and $120$--$240\,$MHz (High Band Antenna; HBA), LOFAR can characterize HzRG spectra down to rest frequencies of $\sim100\,$MHz.  The full complement of stations comprising International LOFAR (I-LOFAR) provides baselines over $1000\,\textrm{km}$, and sub-arcsecond resolution is achievable down to frequencies of about $60\,\textrm{MHz}$. 

At such low radio frequencies, very long baseline interferometry (VLBI) becomes increasingly challenging, as signal propagation through the ionosphere along the different sightlines of widely separated stations gives rise to large differential dispersive delays. These vary rapidly both in time and with direction on the sky, requiring frequent calibration solutions interpolated to the position of the target. Previous works have focused on observations at $\sim150\,$MHz where I-LOFAR is most sensitive and the dispersive delays are less problematic \citep[][Varenius et al., A\&A submitted]{varenius15}. The $\nu^{-2}$ frequency dependence of the ionospheric delays means they are six times larger at 60$\,$MHz than at 150$\,$MHz, reducing the bandwidth over which the assumption can be made that the frequency dependence is linear. Combined with the lower sensitivity of I-LOFAR in the LBA band and the reduction in the number of suitable calibration sources due to absorption processes in compact radio sources below 100$\,$MHz, this makes reducing LBA I-LOFAR observations considerably more challenging than HBA observations. Accordingly, the LBA band of I-LOFAR has been less utilised than the HBA. Previous published LBA results have been limited to observations of 3C 196 \citep{wucknitz10} and the Crab nebula (unpublished) during LOFAR commissioning, when the complement of operational stations limited the longest baseline to $\sim$600$\,$km. 

Here we use I-LOFAR to study the spatially resolved properties of 4C 43.15 (also B3 0731+438) at $z=2.429$ 
\citep{mccarthy91}. This object is one of a sample of 10 that comprise a pilot study of the ultra-steep spectra of HzRGs. We selected 4C 43.15 for this study based on data quality, the suitability of the calibrator, and the simple double-lobed, edge-brightened structure of the target seen at higher frequencies. 
The overall spectral index of 4C 43.15 between 365 MHz \citep[Texas Survey of Radio Sources;][]{texas}  and 1400 MHz \citep[from the Green Bank 1.4 GHz Northern Sky Survey;][]{wb92} is $\alpha=-1.1$, which places it well within the scatter on the $\alpha-z$ relation, seen in Figure~1 of \citet{db00}. 4C 43.15 has been well studied at optical frequencies, and exhibits many of the characteristics of HzRGs \citep[e.g., an extended Lyman-$\alpha$ halo;][]{vm03}. 

Here we present images of 4C 43.15 made with the LBA of I-LOFAR at 55$\,$MHz. These are the first images made with the full operational LBA station complement of I-LOFAR in 2015, and this study sets the record for image resolution at frequencies less than 100$\,$MHz. We compare the low frequency properties of 4C 43.15 with high frequency archival data from the Very Large Array (VLA) to measure the spectral behaviour from 55 -- 4860$\,$MHz. We describe the calibration strategy we designed to address the unique challenges of VLBI for the LBA band of I-LOFAR. The calibration strategy described here provides the foundation for an ongoing pilot survey of ten HzRGs in the Northern Hemisphere with ultra steep ($\alpha<-1$) spectra. 

In \S~\ref{sec:obs} we outline the observations and data pre-processing. Section \ref{sec:lba} describes the LBA calibration, including the VLBI techniques. 
The resulting images are presented in \S~\ref{sec:results} and discussed in \S~\ref{sec:diss}. The conclusions and outlook are summarised in \S~\ref{sec:concl}. Throughout the paper we assume a $\Lambda$CDM concordance cosmology with $H_0=67.8$ km$\,$s$^{-1}\,$Mpc$^{-1}$, $\Omega_{\textrm{m}}=0.308$, and $\Omega_{\Lambda}=0.692$, consistent with \citet{planckcosmo}. At the distance of 4C 43.15, 1\sarc\ corresponds to 8.32$\,$kpc. 

\section{Observations and pre-processing}
\label{sec:obs}
In this section we describe the observations, pre-processing steps and initial flagging of the data. 

As part of project LC3\_018, the target 4C 43.15 was observed on 22 Jan 2015 with 8.5 hr on-source time. Using two beams we conducted the observation with simultaneous continuous frequency coverage between 30 and 78$\,$MHz on both the target and a flux density calibrator. Designed with calibration redundancy in mind, the observation started with 3C 147 as the calibrator and switched to 3C 286 halfway though the observation. Although 3C286 was included for calibrator redundancy, it was later realised that the large uncertainties of the current available beam models prevent accurate calibration transfer to the target at this large angular separation. The observations are summarized in Table~\ref{tab:obs}.

All 46 operational LBA stations participated in the observation, including 24 core stations, 14 remote stations, and 8 international stations. The international stations included 5 in Germany (DE601-DE605) and one each in Sweden (SE607), France (FR606), and the United Kingdom (UK608).  While all stations have 96 dipoles, the core and remote stations are limited by electronics to only using 48 dipoles at one time. The observation was made in the LBA\_OUTER configuration, which uses only the outermost 48 dipoles in the core and remote stations. This configuration reduces the amount of cross-talk between closely spaced dipoles and gives a smaller field of view when compared with other configurations. The international stations always use all 96 dipoles, and thus have roughly twice the sensitivity of core and remote stations. The raw data were recorded with an integration time of 1$\,$s and 64 channels per 0.195$\,$MHz subband to facilitate radio frequency interference (RFI) excision. 

\subsection{Radio Observatory Processing} 
All data were recorded in 8-bit mode and correlated with the COBALT correlator to produce all linear correlation products (XX, XY, YX, YY). After correlation the data were pre-processed by the Radio Observatory. Radio frequency interference was excised using AOFlagger \citep{offringa10} with the default LBA flagging strategy. The data were averaged to 32 channels per subband (to preserve spectral resolution for future studies of carbon radio recombination lines) and 2 second integration time (to preserve information on the time-dependence of phases) before being placed in the Long Term Archive (LTA). The data were retrieved from the LTA and further processed on a parallel cluster kept up to date with the most current stable LOFAR software available at the time (versions 2.9 -- 2.15).

\begin{table*}
\begin{center}
\caption{Observations. The bandwidth for all targets was 48$\,$MHz, split into 244 subbands of 0.195$\,$kHz width. Overlapping times are due to the use of simultaneous beams. Right ascension is hh:mm:ss.ss, and declination is dd:mm:ss.ss.\label{tab:obs}}\vspace{5pt}
\begin{tabular}{llccccccccc}
 & Obs. ID & Object & Type & RA & Dec & Date start & UT start & UT stop & exposure  \\ \hline \\[-5pt]
 & L257205 & 3C 147 & Calibrator & 05:42:36.26 & +49:51:07.08 & 22-Jan-2015 & 18:32:33 & 22:47:32 & 4.25 hr  \\
& L257207 & 4C 43.15 & Target & 07:35:21.89 & +43:44:20.20 & 22-Jan-2015 & 18:32:33 & 22:47:32 & 4.25 hr  \\
& L257209	& 3C 147 & Calibrator &  05:42:36.26 & +49:51:07.08 & 22-Jan-2015 & 22:48:33 & 23:03:32 & 0.25 hr  \\
& L257211	& 3C 286 & Calibrator & 13:31:08.28 & +30:30:32.95 & 22-Jan-2015 & 22:48:33 & 23:03:32 & 0.25 hr  \\
& L257213 & 3C 286 & Calibrator & 13:31:08.28 & +30:30:32.95 & 22-Jan-2015 & 23:04:33 & 03:19:32 & 4.25 hr \\
& L257215 & 4C 43.15 & Target & 07:35:21.89 & +43:44:20.20 & 22-Jan-2015 & 23:04:33 & 03:19:32 & 4.25 hr \\ \hline
\end{tabular}
\end{center}
\end{table*}

\section{Data Calibration}
\label{sec:lba}
\begin{figure*}
	\includegraphics[width=\textwidth]{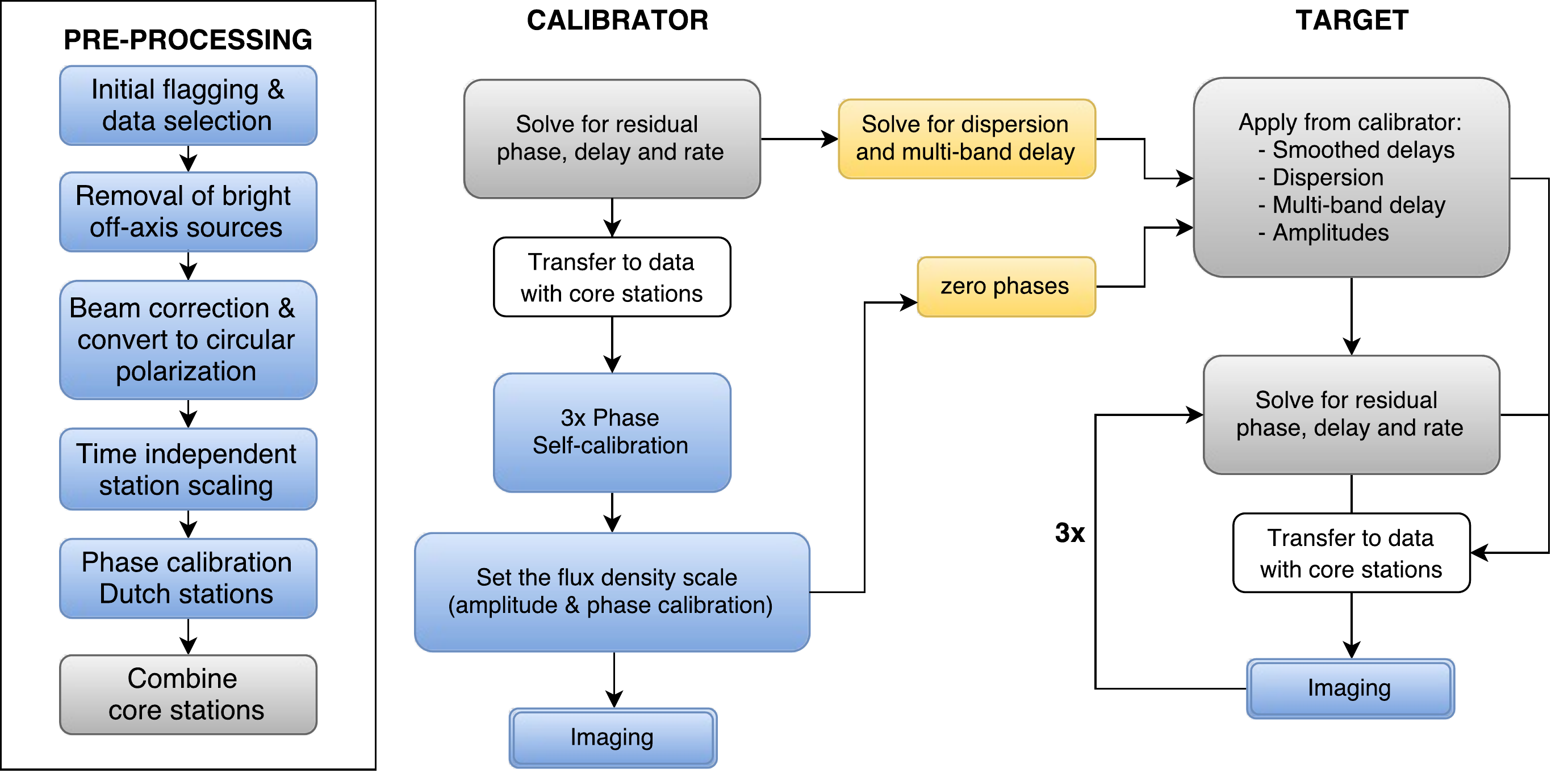}
	\caption{\label{fig:dr} A block diagram overview of the calibration steps. Blue blocks represent operations on data sets with core stations, while gray blocks represent operations on data sets where the core stations have been combined into the `super' station (see \S~\ref{sec:cs} for details on station combination). Yellow blocks represent operations on solution tables rather than data.}
\end{figure*}

In this section we describe in detail the steps taken to calibrate the entire LBA, including international stations, paying particular attention to how we address the unique challenges at low frequencies. Figure~\ref{fig:dr} shows a block diagram overview of the calibration steps.

\subsection{Initial flagging and data selection}
Our first step after downloading the data from the LTA was to run AOFlagger again with the LBA default strategy. Typically 0.5 to 2 per cent of the data in each subband were flagged. An inspection of gain solutions from an initial gain calibration of the entire bandwidth on 3C 147 showed that the Dutch remote station RS409 had dropped out halfway through the first observing block, and we flagged this station and removed it from the dataset.  We further excised one core station (CS501) and one remote station (RS210) after manual inspection. 

We determined the normalised standard deviation per subband from the calibrator data and used this information to select the most sensitive subbands close to the peak sensitivity of the LBA. Outside these subbands the normalised standard deviation rapidly increases towards the edges of the frequency range. The total contiguous bandwidth selected was 15.6$\,$MHz with a central frequency of 55$\,$MHz. During this half of the observation, the standard calibrator 3C 147 was always less than 20 degrees in total angular separation from the target, and the absolute flux density errors are expected to be less than 20 per cent. This is important for two reasons. First, amplitude errors from beam correction models are reduced when objects are close in elevation. The second reason is that we transfer information derived from the calibrator phases (see \S~\ref{sec:tgt} for full details) to the target. This information is valid for a particular direction on the sky, and transferral over very large distances will not improve the signal to noise ratio for the target data. For the second half of the observation, 3C 286 was more than 20 degrees distant from 4C 43.15 for most of that observation block, requiring more advanced calibration which is beyond the scope of this paper, and would only provide $\sqrt{2}$ noise improvement. The second half of the observation was therefore not used for the data analysis in this paper.  

\subsection{Removal of bright off-axis sources}
Bright off-axis sources contribute significantly to the visibilities. At low frequencies, this problem is exacerbated by LOFAR's wide field of view and large primary beam sidelobes. There are several sources that have brightnesses of thousands to tens of thousands of Janskys within the LBA frequency range, and they need to be dealt with. We accomplished the removal of bright off-axis sources using a method called demixing \citep{vdt07}, where the data are phase-shifted to the off-axis source, averaged to mimic beam and time smearing, and calibrated against a model. All baselines were demixed, although simulations performed as part of commissioning work showed that the source models have insufficient resolution to correctly predict the compact bright sources to which the longest baselines would be sensitive. Such sources produce strong beating in the amplitudes of the visibilities, which is visible by eye. A careful visual inspection ensured that this was not a problem for these data. Using the calibration solutions, the \textit{uncorrected} visibilities for the source are subtracted. After examination of the bright off-axis sources above the horizon and within $90^{\circ}$ of the target and calibrator (such a large radius is necessary in case there are sources in sidelobes), we demixed Cassiopeia A and Taurus A from our data. After demixing the data were averaged to 16 channels per subband to reduce the data volume, and the AOFlagger was run again with the default LBA flagging strategy. Typical flagging percentages were 2--4 per cent. The combined losses from time (2$\,$sec) and bandwidth (4 channels per 195$\,$kHz subband) smearing on the longest baseline are 5 per cent at a radius of 95 arcsec \citep{bridleschwab}. Higher frequency observations of 4C 43.15 show its largest angular size to be 11 arcsec, well within this field of view. 

\subsection{LOFAR beam correction and conversion to circular polarization}
At low frequencies, differential Faraday rotation from propagation through the ionosphere can shift flux density from the XX and YY to the cross hand polarizations. An effective way to deal with this is to convert from linear to circular polarization, which shifts the impact of differential Faraday rotation to only a L-R phase offset in the resulting circular polarization. Since the conversion from linear to circular polarization is beam dependent, we first removed the beam. 
We used \textsc{mscorpol} (version 1.7)\footnote{\textsc{mscorpol} was developed by T.~D. Carozzi and available at: \\ \href{https://github.com/2baOrNot2ba/mscorpol}{\color{blue}{https://github.com/2baOrNot2ba/mscorpol}}} to accomplish both removal of beam effects and conversion to circular polarization. This software performs a correction for the geometric projection of the incident electric field onto the antennas, which are modelled as ideal electric dipoles. 
One drawback of \textsc{mscorpol} is that it does not yet include frequency dependence in the beam model, so we also replicated our entire calibration strategy but correcting for the beam with the LOFAR \emph{new default pre-processing pipeline} (NDPPP), which has frequency-dependent beam models, rather than \textsc{mscorpol}. We converted the NDPPP beam-corrected data to circular polarization using standard equations, and followed the same calibration steps described below. We found that data where the beam was removed with \textsc{mscorpol} ultimately had more robust calibration solutions and better reproduced the input model for the calibrator. Therefore we chose to use the \textsc{mscorpol} beam correction. 

\subsection{Time-independent station scaling}
The visibilities for the international stations must be scaled to approximately the right amplitudes relative to the core and remote stations before calibration. This is important because the amplitudes of the visibilities are later used to calculate the data weights, which are used in subsequent calibration steps, see \S~\ref{sec:pdr}. To do this we solved for the diagonal gains (RR,LL) on all baselines using the Statistical Efficient Calibration \citep[StEfCal;][]{sw14} algorithm in NDPPP. 
 One solution was calculated every eight seconds per 0.915$\,$MHz bandwidth (one subband). The StEfCal algorithm calculates time and frequency independent phase errors, and does not take into account how phase changes with frequency (the delay; $d\phi/d\nu$) or time (the rate; $d\phi/dt$). If the solution interval over which StEfCal operates is large compared to these effects, the resulting incoherent averaging will result in a reduction in signal to noise. Since the incoherently averaged amplitudes are adjusted to the correct level, the coherence losses manifest as an increase of the noise level.  
 Using the maximal values for delays and rates found in \S~\ref{sec:pdr} to calculate the signal to noise reduction \citep[from Eqn. 9.8 and 9.11 of ][]{moran95}, we find losses of 6 and 16 per cent for delays and rates, respectively. 
 
 The calibrator 3C 147 flux density was given by the model from \citet{sh12}. 
3C 147 is expected to be unresolved or only marginally resolved and therefore expeted to provide an equal amplitude response to baselines of any length. 
We use this gain calibration for two tasks: (i) to find an overall scaling factor for each station that correctly provides the relative amplitudes of all stations; and (ii) to identify bad data using the LOFAR Solution Tool\footnote{The LOFAR Solution Tool (LoSoTo) was developed by Francesco de Gasperin and is available at: \\ \href{https://github.com/revoltek/losoto}{\color{blue}{https://github.com/revoltek/losoto}}}. About 20 per cent of the solutions were flagged either due to outliers or periods of time with loss of phase coherence, and we transferred these flags back to the data. 
To find the time-independent scaling factor per station, we zeroed the phases and calculated a single time-averaged amplitude correction for each antenna. These corrections were applied to both calibrator and target datasets. 

\subsection{Phase calibration for Dutch stations}
 We solved for overall phase corrections using only the Dutch array but filtering core -- core station baselines, which can have substantial low-level RFI and are sensitive to extended emission. The phase calibration removes ionospheric distortions in the direction of the dominant source at the pointing centre. We performed the phase calibration separately for 3C 147 and 4C 43.15 using appropriate skymodels. 3C 147 is the dominant source in its field, and we use the \citet{sh12} point source model. 4C 43.15 has a flux density of at least 10$\,$Jy in the LBA frequency range. We used an apparent sky model of the field constructed from the TGSS Alternative Data Release 1 \citep{tgssadr}, containing all sources within 7 degrees of our target and with a flux density above 1$\,$Jy. 

\subsection{Combining core stations}
\label{sec:cs}
After phase calibration of the Dutch stations for both the calibrator and the target, we coherently added the visibilities from the core stations to create a `super' station. 
This provides an extremely sensitive `super' station with increased signal to noise on individual baselines to anchor the I-LOFAR calibration (described further in \S~\ref{sec:pdr}). 
All core stations are referred to a single clock and hence should have delays and rates that are negligibly different after phase calibration is performed. The station combination was accomplished with the Station Adder in NDPPP by taking the weighted average of all visibilities on particular baselines. For each remote and international station, all visibilities on baselines between that station and the core stations are averaged together taking the data weights into account. The new $u,v,w$ coordinates are calculated as the weighted geometric center of the $u,v,w$ coordinates of the visibilities being combined\footnote{We found an extra 1 per cent reduction in noise for the calibrator when using the weighted geometric center of the $u,v,w$ coordinates, rather than calculating the $u,v,w$ coordinates based on the `super' station position. This has been implemented in NDPPP (LOFAR software version 12.2.0)}. Once the core stations were combined, we created a new data set containing only the `super' station and remote and international stations. The dataset with the uncombined core stations was kept for later use. 
The final averaging parameters for the data were 4 channels per subband for 3C 147, and 8 channels per subband for 4C 43.15. 
After averaging the data were again flagged with the AOFlagger default LBA flagging strategy, which flagged another 1 -- 2 per cent of the data.

\subsection{Calibrator residual phase, delay, and rate}
\label{sec:pdr}
The international stations are separated by up to 1292 km and have independent clocks which time stamp the data at the correlator. There are residual non-dispersive delays due to the offset of the separate rubidium clocks at each station. Correlator model errors can also introduce residual non-dispersive delays up to $\sim100$ns. 
Dispersive delays from the ionosphere make a large contribution to the phase errors. Given enough signal to noise on every baseline, we could solve for the phase errors over small enough time and bandwidth intervals that the dispersive errors can be approximated as constant. However, a single international-international baseline is only sensitive to sources of $\sim10$Jy over the resolution of our data ($\Delta\nu=0.195\,$MHz, 2 sec). Larger bandwidth and time intervals increase the signal to noise ratio, and the next step is to model the dispersive delays and rates with linear slopes in frequency and time. 
This can be done using a technique known as fringe-fitting \citep[e.g.,][]{cotton95,tms01}. A global fringe-fitting algorithm is implemented as the task FRING in the Astronomical Image Processing System \citep[AIPS;][]{greisen03}. We therefore converted our data from measurement set to UVFITS format using the task \textsc{ms2uvfits} and read it into AIPS. The data weights of each visibility were set to be the inverse square of the standard deviation of the data within a three minute window. 

The ionosphere introduces a dispersive delay, where the phase corruption from the ionosphere is inversely proportional to frequency, $\phi_{\textrm{ion}} \propto \nu^{-1}$. The dispersive delay is therefore inversely proportional to frequency squared, $d\phi / d\nu \propto -\nu^{-2}$. Non-dispersive delays such as those introduced by clock offsets are frequency-independent. The ionospheric delay is by far the dominant effect. For a more in-depth discussion of all the different contributions to the delay at 150$\,$MHz for LOFAR, see \citet{moldon15}. The delay fitting-task FRING in AIPS fits a single, non-dispersive delay solution to each so-called \emph{intermediate frequency} (IF), where an IF is a continuous bandwidth segment. With I-LOFAR data, we have the freedom to choose the desired IF bandwidth by combining any number of LOFAR subbands (each of width 0.195$\,$MHz). This allows us to make a piece-wise linear approximation to the true phase behaviour. Making wider IFs provides a higher peak sensitivity, but leads to increasingly large deviations between the (non-dispersive only) model and the (dispersive and non-dispersive) reality at the IF edges when the dispersive delay contribution is large. As a compromise, we create 8 IFs of width 1.95$\,$MHz each (10 LOFAR subbands), and each IF is calibrated independently. We used high resolution model of 3C 147 (from a previous I-LOFAR HBA observation at 150$\,$MHz) for the calibration, and set the total flux density scale from \citet{sh12}. The solution interval was set to 30 seconds, and we found solutions for all antennas using only baselines with a projected separation $>10$k$\lambda$, effectively removing data from all baselines containing only Dutch stations. The calibration used the `super' station as the reference antenna. 

The search windows were limited to 5$\,\mu$s for delays and 80$\,$mHz for rates. Typical delays for remote stations were 30$\,$ns, while international station delays ranged from 100$\,$ns  to 1$\,\mu$s. The delay solutions showed the expected behaviour, with larger offsets from zero for longer baselines, and increasing magnitudes (away from zero) with decreasing frequency. Rates were typically 
up to a few tens of mHz for remote and international stations. 

\subsection{Calibrator phase self-calibration}
The combined `super' station, while useful for gaining signal to noise on individual baselines during fringe-fitting, left undesirable artefacts when imaging. This can occur if the phase-only calibration prior to station combination is imperfect. The imperfect calibration will result in the `super' station not having a sensitivity equal to the sum of the constituent core stations. The `super' station also has a much smaller field of view than the other stations in the array. Therefore we transferred the fringe-fitting solutions to a dataset where the core stations were not combined. 

 Before applying the calibration solutions we smoothed the delays and rates with solution intervals of 6 and 12 minutes, respectively, after clipping outliers (solutions more than 20 mHz and 50 ns different from the smoothed value within a 30 minute window for rates and delays, respectively). The smoothing intervals were determined by comparing with the unsmoothed solutions to find the smallest time window that did not oversmooth the data. We applied the solutions to a dataset where the core stations were not combined. 
The data were then averaged by a factor of two in time prior to self-calibration to 4 second integration times. We performed three phase-only self-calibration loops with time intervals of 30 seconds, 8 seconds, and 4 seconds. Further self-calibration did not improve the image fidelity or reduce the image noise. 

\subsection{Setting the flux density scale}
\begin{figure}
\includegraphics[width=0.5\textwidth]{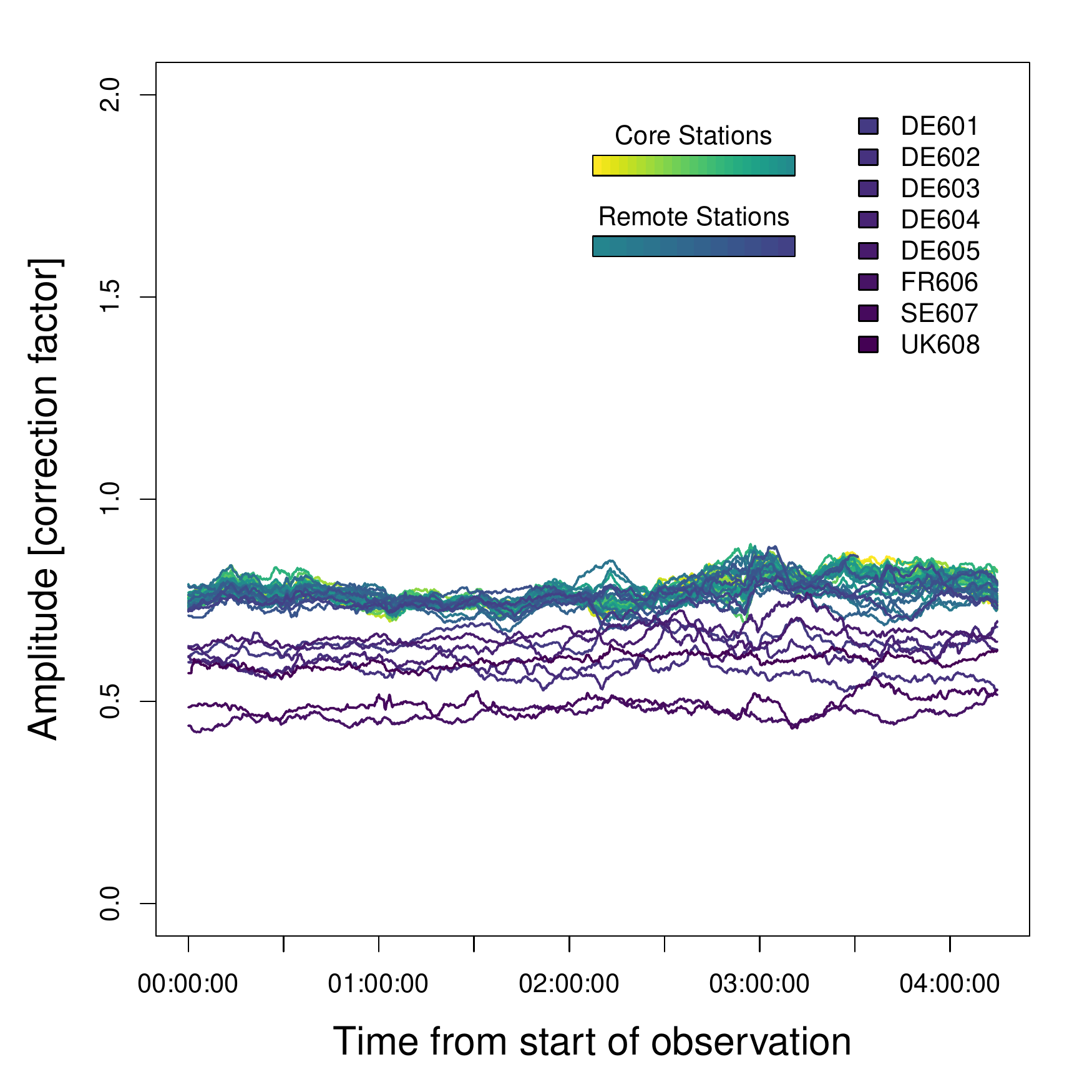}
\caption{Amplitude solutions for all stations, from the final step of self-calibration. These are \textit{corrections} to the initial amplitude calibration of each station, for the central IF at 53$\,$MHz. The colours go from core stations (darkest) to international stations (lightest). \label{fig:amp}}
\end{figure}

After applying the final phase-only calibration, we solved for amplitude and phase with a 5 minute solution interval, as the amplitudes vary slowly with time. The amplitude solutions provide time-variable \textit{corrections} to the initial default station amplitude calibration. Fig.~\ref{fig:amp} shows the amplitude solutions per station for an IF near the centre of the band. 

The amplitude solutions show some small-scale variations in time, but are stable to within 20 per cent of the median value over the entirety of the observation. We therefore adopt errors of 20 per cent for the measurements presented here. Several effects could be responsible for the variations in time such as imperfect beam or source models, or ionospheric disturbances. Currently we are not able at this time to distinguish whether the time variation we see is from the ionosphere or beam errors.  

\begin{figure}
\includegraphics[width=0.5\textwidth]{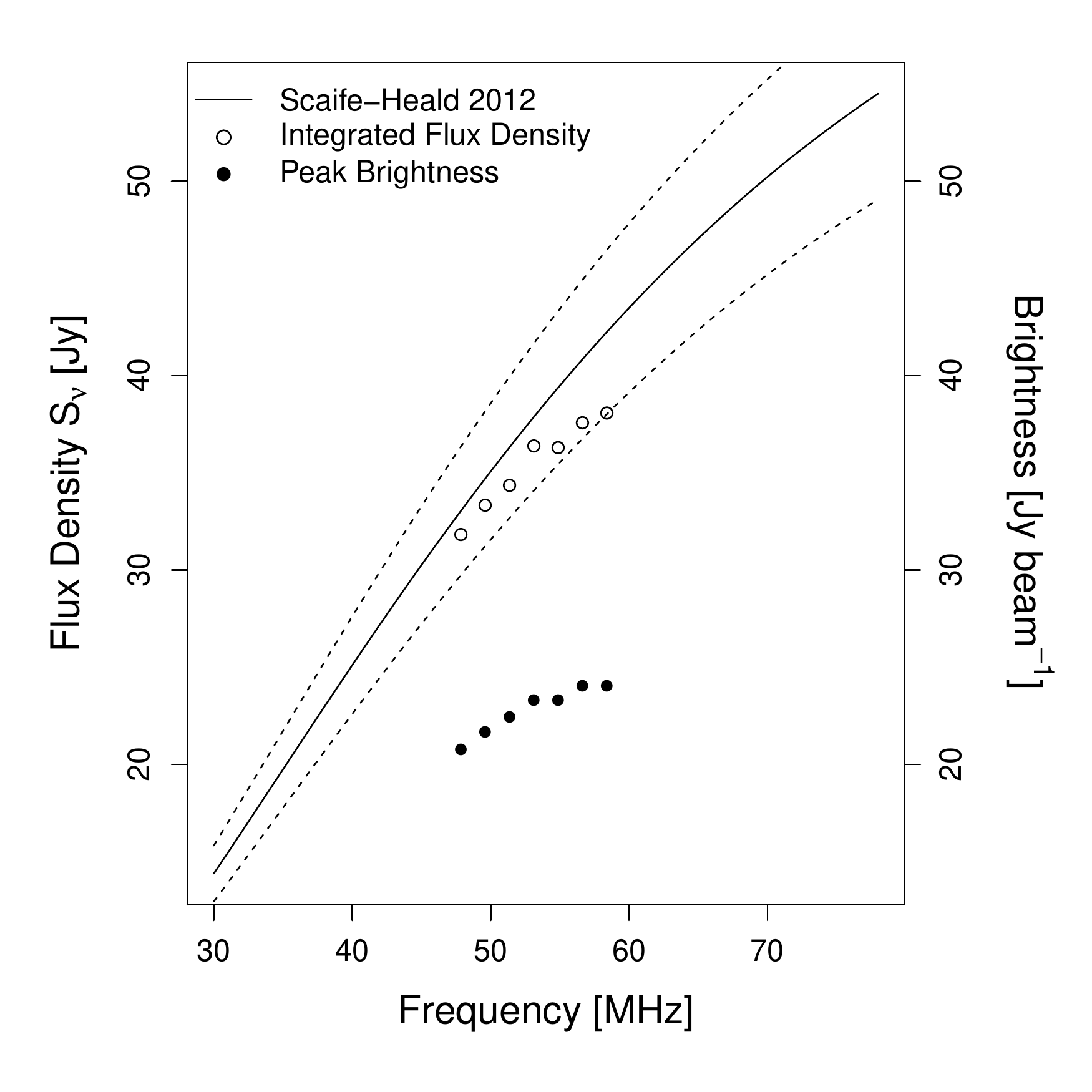}
\caption{Spectral energy distribution (SED) of 3C 147. The solid line shows the calibration model that we used, while the dashed lines indicate errors of 10 per cent. The open circles show the integrated flux density measurements from each IF, while the filled circles show the peak brightness measurements. \label{fig:calsed} }
\end{figure}

\begin{figure}
\includegraphics[width=0.5\textwidth]{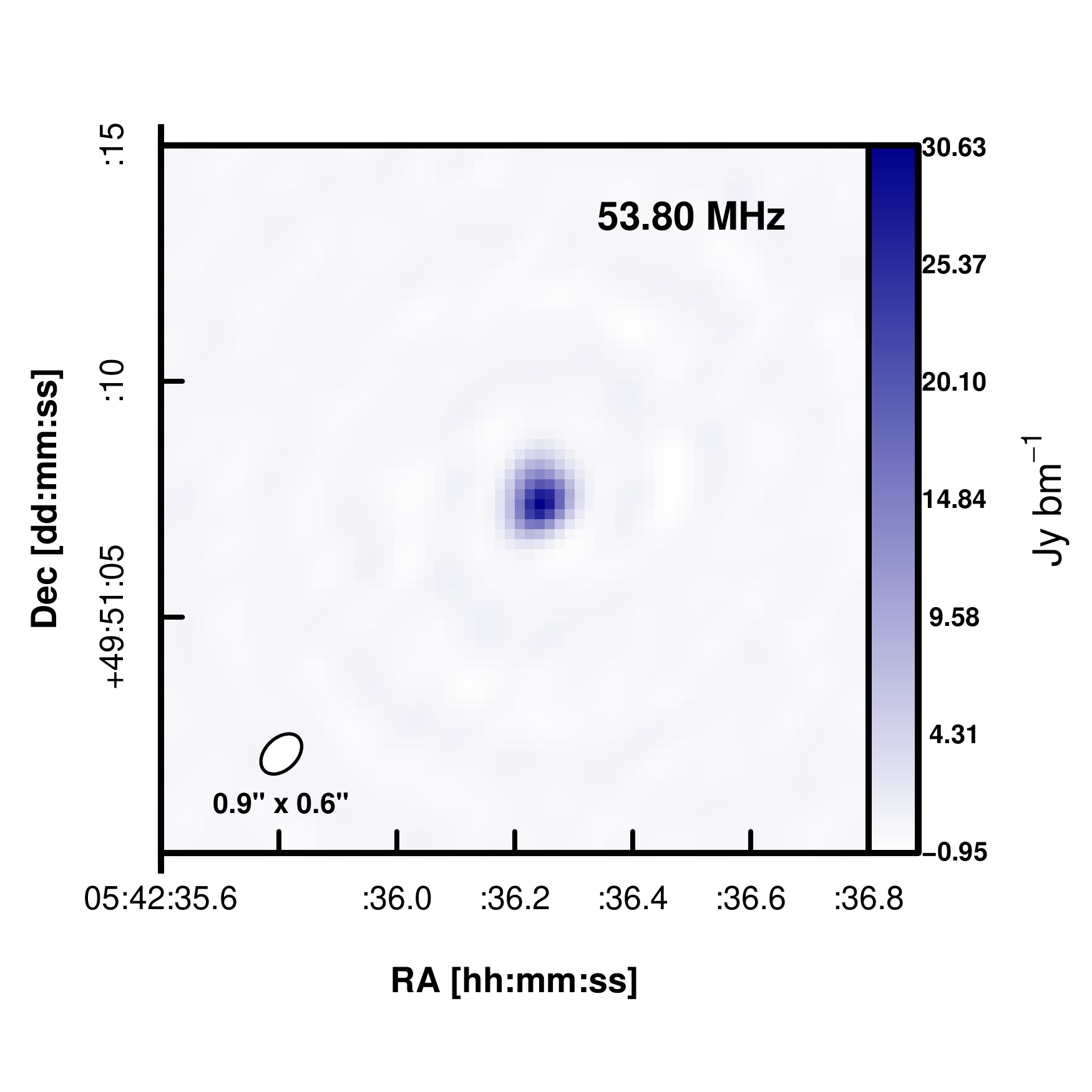}
\caption{The calibrator, 3C 147, imaged using 15.6$\,$MHz bandwidth and 4.25 hr of data. The noise in the image is 135$\,$mJy$\,$bm$^{-1}$. The NW-elongation is a jet also seen at higher frequencies. \label{fig:cal}}
\end{figure}

We checked the calibration of 3C 147 by imaging each IF of the final self-calibrated data separately, fitting a Gaussian to extract the integrated flux density, and plotting this against the input model, see Figure~\ref{fig:calsed}. The integrated flux density measurements are within the errors of the point-source model, while the peak brightness measurements are below the model. This is due to the fact that the jet in 3C 147, which is seen also at higher frequencies, is resolved (the restoring beam is 0.9\sarc\ $\times$ 0.6\sarc\ ). The values are systematically lower than the model, and slightly flatter. This could be due to the fact that the starting model from \cite{sh12} is a point source model, and 3C 147 is resolved. The flattening spectral index towards higher frequencies, where the beam size is smaller, implies that the jet which appears as a NW-elongation in Figure~\ref{fig:cal} has a steeper low-frequency  spectral index than the core. This is supported by the fact that the peak brightness measurements are slightly flatter than the integrated flux density measurements in Figure~\ref{fig:calsed}.  

In some extremely compact objects, scintillation effects from the interstellar medium have been seen to artificially broaden sources \citep[e.g.,][]{linsky08,quirrenbach92,rickett86}. However, these scintillations are usually only seen in compact ($\sim$10 mas) sources and/or on longer timescales (days to weeks). Both the calibrator and target are larger in size, and well outside of the galactic plane (above $b=20^{\circ}$). We thus do not expect that they should be impacted. The final self-calibrated image using the entire bandwidth is shown in Fig.~\ref{fig:cal}, and has a noise of 135$\,$mJy$\,$bm$^{-1}$, about a factor of 3 above the expected noise given the amount of flagging (40 per cent) and the $u-v$ cut in imaging ($>20\,$k$\lambda$).  

\subsection{Target residual phase, delay, and rate}
\label{sec:tgt}
Before fringe-fitting on the target, the time independent and dependent amplitude corrections derived from the calibrator were applied to the target, for a dataset with the `super' station. The time dependent core station amplitude corrections were all within a few per cent of each other so we transferred the amplitude corrections from a station close to the centre of the array, CS001, to the `super' station. The fringe-fitting solutions from the calibrator, approximately $20^{\circ}$ away, should also contain some instrumental and systematic effects, such as those due to clock offsets and large-scale ionospheric structure, which will be common to the target direction and can be usefully subtracted by applying the calibrator solutions to the target data. 
After extensive testing, we found that we gained the most signal to noise in the fringe-fitting by applying the smoothed delays from the calibrator, along with a model of the frequency dependence of the phases. We used the AIPS task MBDLY to model the frequency dependence from the FRING calibration solutions with smoothed delays. We used the `DISP' option of MBDLY to find the dispersion and multi-band delay for each solution in the fringe-fitting calibration table. After zeroing the phases and rates in the FRING calibration solutions, we used the MBDLY results to correct for the multi-band delay and the dispersion. With the phases already zeroed, the dispersion provides a \emph{relative} correction of the phases, effectively removing the frequency dependence. 
This allowed us to use a wider bandwidth in the FRING algorithm, which increased the signal to noise. We chose to use the entire 15.6$\,$MHz bandwidth. 
The resultant delays were smaller by at least a factor of two on the longest baselines, which was expected as transferring the delays from the calibrator already should have corrected the bulk of the delays. These residual delays are then the \emph{difference} in the dispersion and multi-band delays between the target and the calibrator. We also tested 
the effect of only including data from partial $uv$ selections and established that it was necessary to use the full $uv$ range to find robust fringe-fitting solutions. It is important to remember that the shortest baseline is from the `super' station to the nearest remote station. There are 12 remote station -- `super' station baselines, ranging from about 4$\,$km to 55$\,$km, with a median length of about 16$\,$km. 

The next step was to perform fringe fitting on the target. We began fringe fitting using a point source model with a flux density equal to the integrated flux density of the target measured from a low-resolution image made with only the Dutch array. Initial tests showed a double source with similar separation and position angle (PA) as seen for 4C 43.15 at higher frequencies, rather than the input point source model. We further self-calibrated by using the resulting image as a starting model for fringe-fitting. We repeated this self-calibration until the image stopped improving.   

\subsection{Astrometric Corrections}
The process of fringe frequency fitting does not derive absolute phases or preserve absolute positions, only relative ones. To derive the absolute astrometric positions we assumed that the components visible in our derived images coincided with the components visible on the high-frequency archival data for which the absolute astrometry was correct. 
We centred the low-frequency lobes in the 
direction perpendicular to the jet axis, and along the jet axis we centred the maximum extent of the low-frequency emission between the maximum extent of the high frequency emission. The re-positioning of the source is accurate to within $\sim$0.6\sarc\ assuming that the total extent of the low-frequency emission is contained within the total extent of the high-frequency emission. This positional uncertainty will not affect the following analysis. 

\section{Results}
\label{sec:results}

\begin{figure*} 
\includegraphics[width=0.49\textwidth,clip,trim=1.8cm 1cm 1.8cm 0cm]{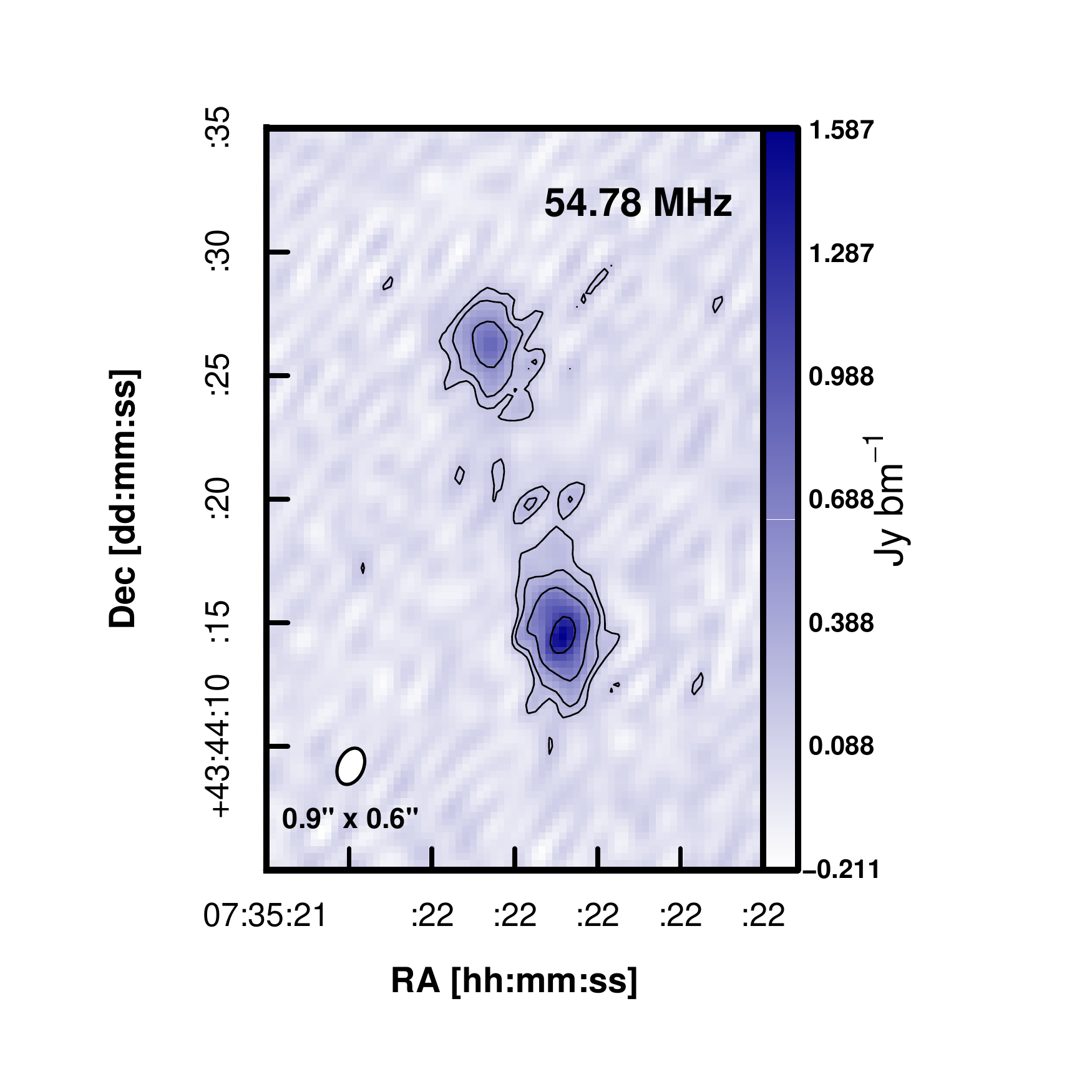} 
\includegraphics[width=0.49\textwidth,clip,trim=1.8cm 1cm 1.8cm 0cm]{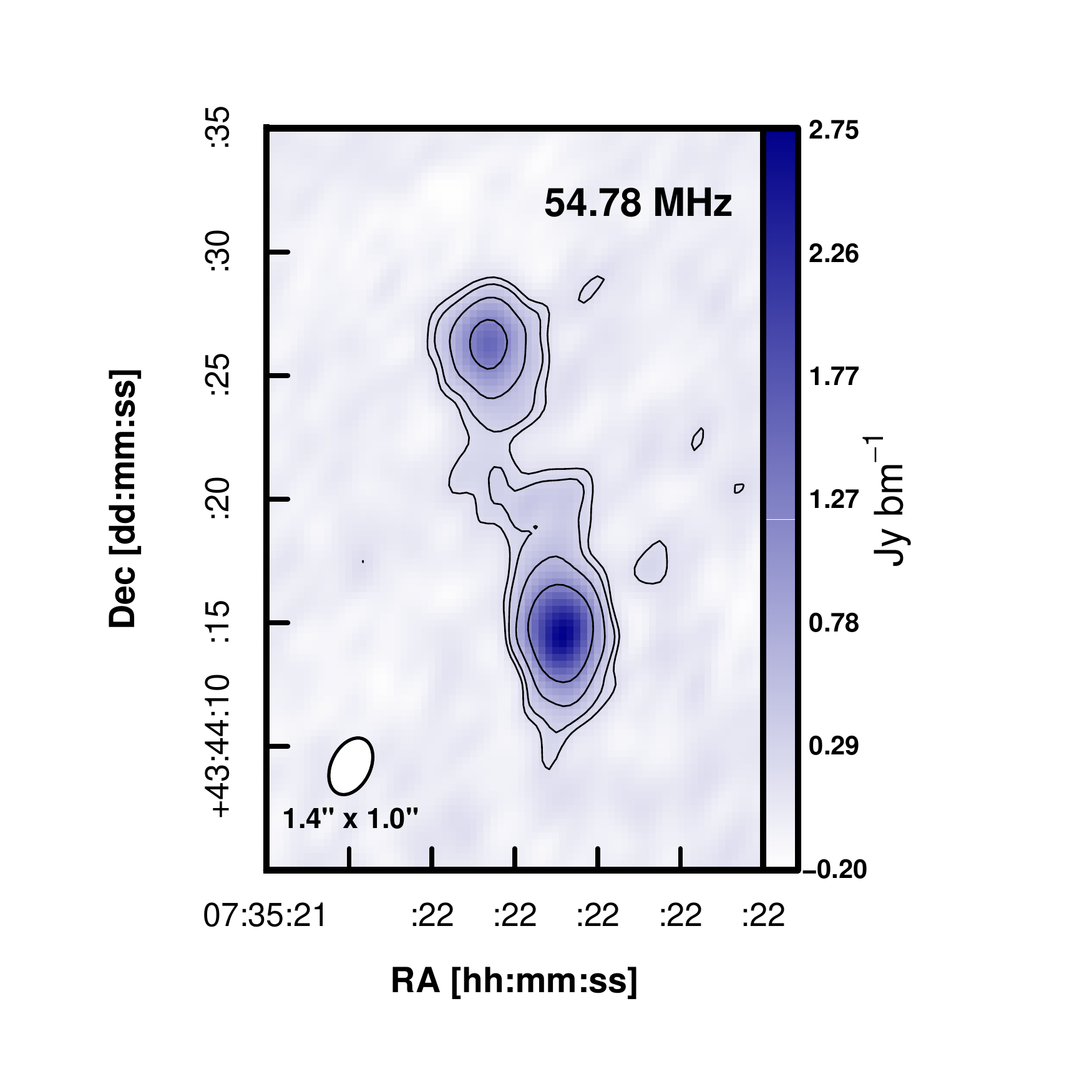} 
\caption{\label{fig:tgt} The final LBA images of 4C 43.15. The image on the left was made using 15.6$\,$MHz of bandwidth centred on 55$\,$MHz. We used the multi-scale function of the \textsc{CLEAN} task in CASA with Briggs weighting (robust -1.5) and no inner $uv$ cut. The image noise achieved is 59\mjy\ , while the expected noise given the amount of flagged data and image weighting is 25\mjy\ . The final restoring beam is 0.9\sarc\ $\times$0.6\sarc\ with PA -33 deg. The image on the right is the same image, but smoothed with a Gaussian kernel 1.2 times the size of the restoring beam. The contours in both images are drawn at the same levels, which are 3, 5, 10, and 20$\sigma$ of the unsmoothed image. }
\end{figure*}

In Figure~\ref{fig:tgt} we present an LBA image of 4C 43.15 which achieves a resolution of 0.9\sarc\ $\times$0.6\sarc\ with PA -33 deg and has a noise level of 59\mjy\ . This image was made using multi-scale CLEAN in the Common Astronomy Software Applications \citep[CASA;][]{casa} software package, with Briggs weighting and a robust parameter of -1.5, which is close to uniform weighting and offers higher resolution than natural weighting. The contours show the significance of the detection (starting at $3\sigma$ and up to $20\sigma$). \emph{This is the first image made with sub-arcsecond resolution at frequencies below 100$\,$MHz.} The beam area is a factor of 2.5 smaller than that achieved by \citet{wucknitz10}. The measured noise is a factor of 2.4 above the theoretical noise. 

In the following subsections we examine first the morphology of 4C 43.15 and then the spectral index properties of the source. For comparison with higher frequencies, we used archival data from the NRAO VLA Archive Survey\footnote{http://archive.nrao.edu/nvas/}. The available images had higher resolution than the LBA image presented here, with the exception of images at 1.4$\,$GHz. We therefore downloaded and re-imaged the calibrated data to produce more similar beam sizes with the use of different weighting and/or maximum baseline length. The archival data and resulting beam sizes are listed in Table~\ref{tab:vla}. All images were then convolved to the largest beam full width at half maximum (FWHM) of 1.55\sarc\ $\times$ 0.98\sarc\ (at 1.4$\,$GHz). Even at the highest frequency used here (8.4$\,$GHz) the A-configuration of the VLA is still sensitive to emission on scales of about 5\sarc\ , or roughly the size of a single lobe of 4C 43.15. We therefore do not expect that the image misses significant contributions to the flux density. This is supported by the third panel in Figure~\ref{fig:si}, which shows that the spectral indices from 1.4$\,$GHz to the two higher frequencies in this study are the same within the errors. If a substantial amount of flux density were missing at 8.4$\,$GHz, we would expect to see a steeper spectral index from 1.4$\,$GHz to 8.4$\,$GHz than from 1.4$\,$GHz to 4.7$\,$GHz.  

\begin{table}
\caption{\label{tab:vla} Summary of archival VLA data and re-imaging parameters. All data were taken in A-configuration, which has a minimum baseline of 0.68$\,$km and a maximum baseline of 36.4$\,$km.}
\begin{tabular}{ccccc}
Date & $\nu$ &  & Maximum & Beam  \\[-2pt]
observed & $[$GHz$]$ &  Weighting & baseline  & size  \\ \hline
31-08-1995 & 1.4 &  super uniform & -- &  1.55\sarc\ $\times$ 0.98\sarc\  \\
19-03-1994	& 4.7 & natural & 192$\,$k$\lambda$ & 1.02\sarc\ $\times$ 0.88\sarc  \\
31-08-1995 & 8.4 &  natural & 192$\,$k$\lambda$ & 1.05\sarc\ $\times$ 0.83\sarc\  \\
\end{tabular}
\end{table}

\subsection{Morphology}
Figure~\ref{fig:tgt} shows two radio lobes that are edge brightened, the classic signature of an FR$\,$II source. FR$\,$II sources have several components. There are collimated jets that extend in opposite directions from the host galaxy, terminating in hot spots that are bright, concentrated regions of emission. More diffuse, extended radio emission from plasma flowing back from the hot spots comprises the lobes. In HzRGs, only the hotspots and lobes have been directly observed, since the jets have low surface brightness. Observations of 4C 43.15 at frequencies higher than 1$\,$GHz clearly show the hot spots and diffuse lobe emission, but this is the first time this morphology has been spatially resolved for an HzRG at frequencies $<300\,$MHz. The smoothed image shows a bridge of emission connecting the two lobes at the 3 and 5$\sigma$ significance levels. This is
similar to what is seen in the canonical low-redshift FR$\,$II galaxy, Cygnus A \citep{carilli91}, but this is \emph{the first time that a bridge of low frequency emission connecting the two lobes has been observed in a HzRG.}

\begin{figure}
\includegraphics[width=0.5\textwidth,clip,trim=1cm 0.5cm 4cm 4cm]{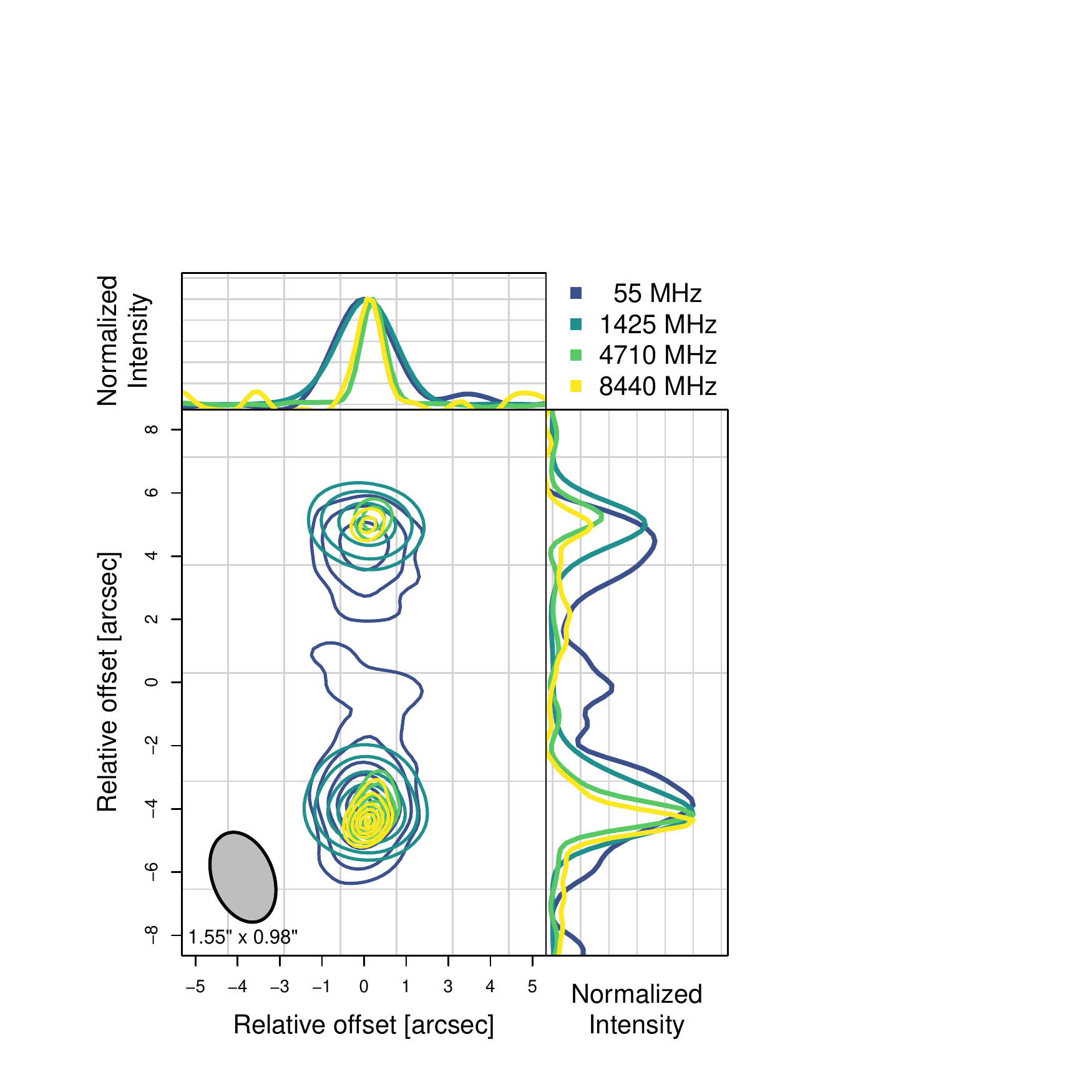}
\caption{\label{fig:rot} Contours and intensity profiles for 4C 43.15 at four frequencies.  The rotation angle of the jet was determined per frequency to rotate all images so the jet axis is aligned for all images. The contours are set at 20, 40, 60, 80, and 95 per cent of the maximum intensity (which is unity).}
\end{figure}

To qualitatively study the low-frequency morphology of 4C 43.15 in more detail and compare it with the structure at high frequencies, we derived the brightness profiles along and perpendicular to the source axis. To do this we defined the jet axis by drawing a line between the centroids of Gaussian fits to each lobe. We used the position angle of this line to rotate all images (the unsmoothed image was used for the 55$\,$MHz image) so the jet axis is aligned with North. We fitted for the rotation angle independently for all frequencies, and found the measured position angles were all within 1 degree of each other, so we used the average value of 13.36 degrees to rotate all images. The rotated images are shown overlaid on each other in Figure~\ref{fig:rot}, along with normalized sums of the flux density along the North--South direction and East--West direction. 

The integrated flux density ratio of the lobes also evolves with frequency, which can be seen in Figure~\ref{fig:rot}. The lobe ratio changes from 3 at the highest frequency to 1.7 at the lowest frequency. This implies a difference in spectral index between the two lobes, which will be discussed in the next section.

\subsection{Spectral Index Properties}
\label{sec:si}

\begin{figure}
\includegraphics[width=0.5\textwidth]{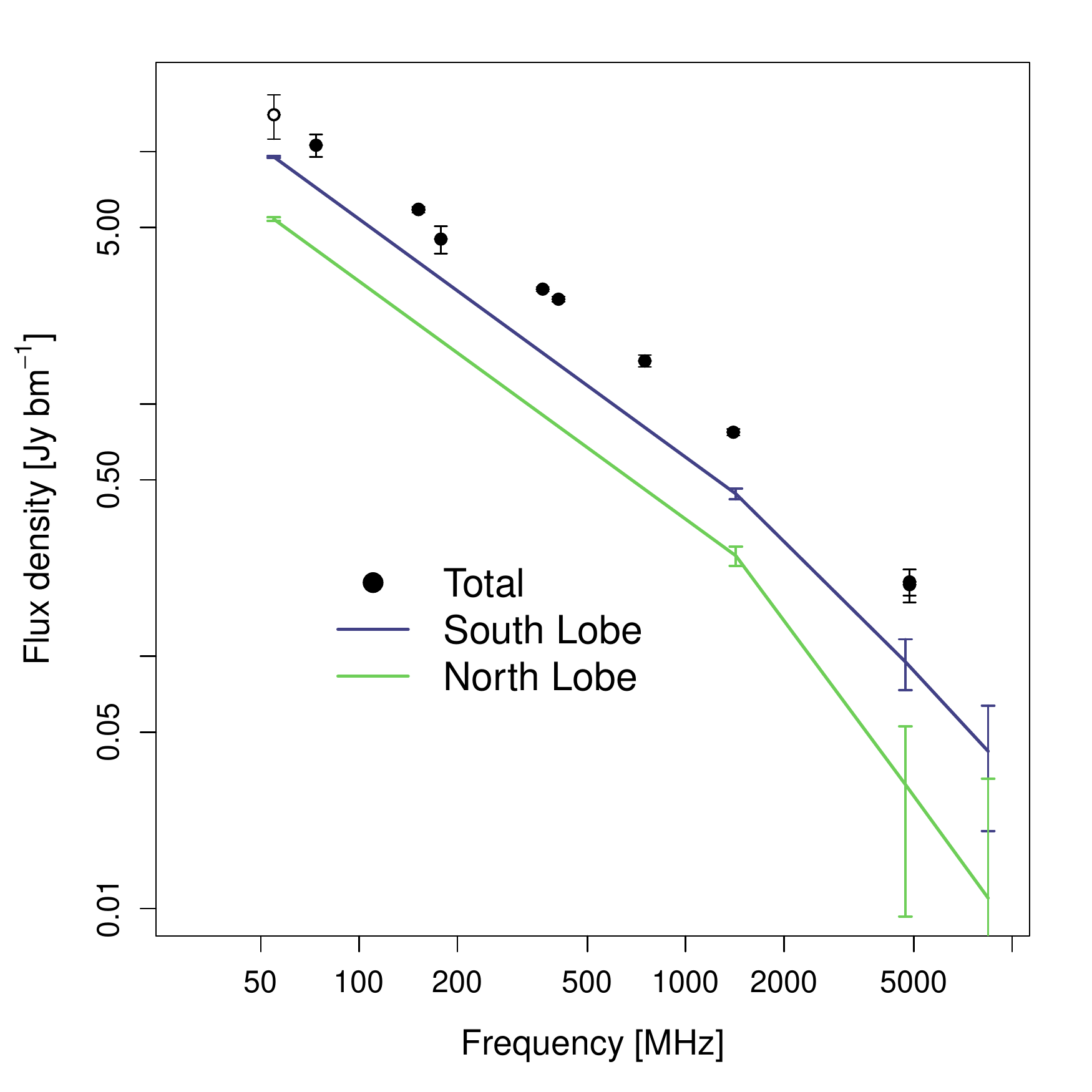}
\caption{\label{fig:bb} The total integrated spectrum derived from archival (black circles) and LOFAR data (white circle with black outline). The integrated spectra of the lobes are also shown for the measurements described in \S~\ref{sec:si}. The lines between data points do not represent fits to the data and are only drawn to guide the eye.}
\end{figure}

In this section we shall describe the spectral index properties of 4C 43.15 using the integrated spectra from each of the lobes, and the total integrated spectral index.  Figure~\ref{fig:bb} shows the lobe spectra and the total integrated spectrum for comparison. The lobe spectra at 1.4, 4.7, and 8.4$\,$GHz were measured from VLA archival images convolved to the resolution at 1.4$\,$GHz and are reported in Table~\ref{tab:sp}. We assumed errors of 20 per cent for the LOFAR data and 5 per cent for the VLA archival data.  The integrated spectral data were taken from the NASA/IPAC Extragalactic Database (NED), with the inclusion of the new LOFAR data point, see Table~\ref{tab:ned}. 

\begin{table}
\caption{\label{tab:ned}Integrated Flux Density Measurements.}
\begin{tabular}{lccl}
Frequency & Flux Density & Error & Reference \\
 & [Jy] & [Jy] &  \\ \hline
54$\,$MHz & 14.9 & 3.0  & This work \\
74$\,$MHz & 10.6 & 1.1 & VLSS, \citet{cohen07} \\
151$\,$MHz & 5.9 & 0.17 & 6C, \citet{6c} \\
178$\,$MHz & 4.5 & 0.56 & 3C, \citet{gower67} \\
365$\,$MHz & 2.9 & 0.056 &  Texas, \citet{texas} \\
408$\,$MHz & 2.6 & 0.056 & Bologna, \citet{bologna} \\
750$\,$MHz & 1.5 & 0.080 & \citet{pt66} \\
1.4$\,$GHz & 0.77 & 0.023 & NVSS, \citet{nvss} \\
4.85$\,$GHz & 0.19 & 0.029 & \citet{becker91} \\
\end{tabular}
\end{table}

\begin{figure}
\includegraphics[width=0.5\textwidth,clip,trim=3cm 2cm 3cm 2cm]{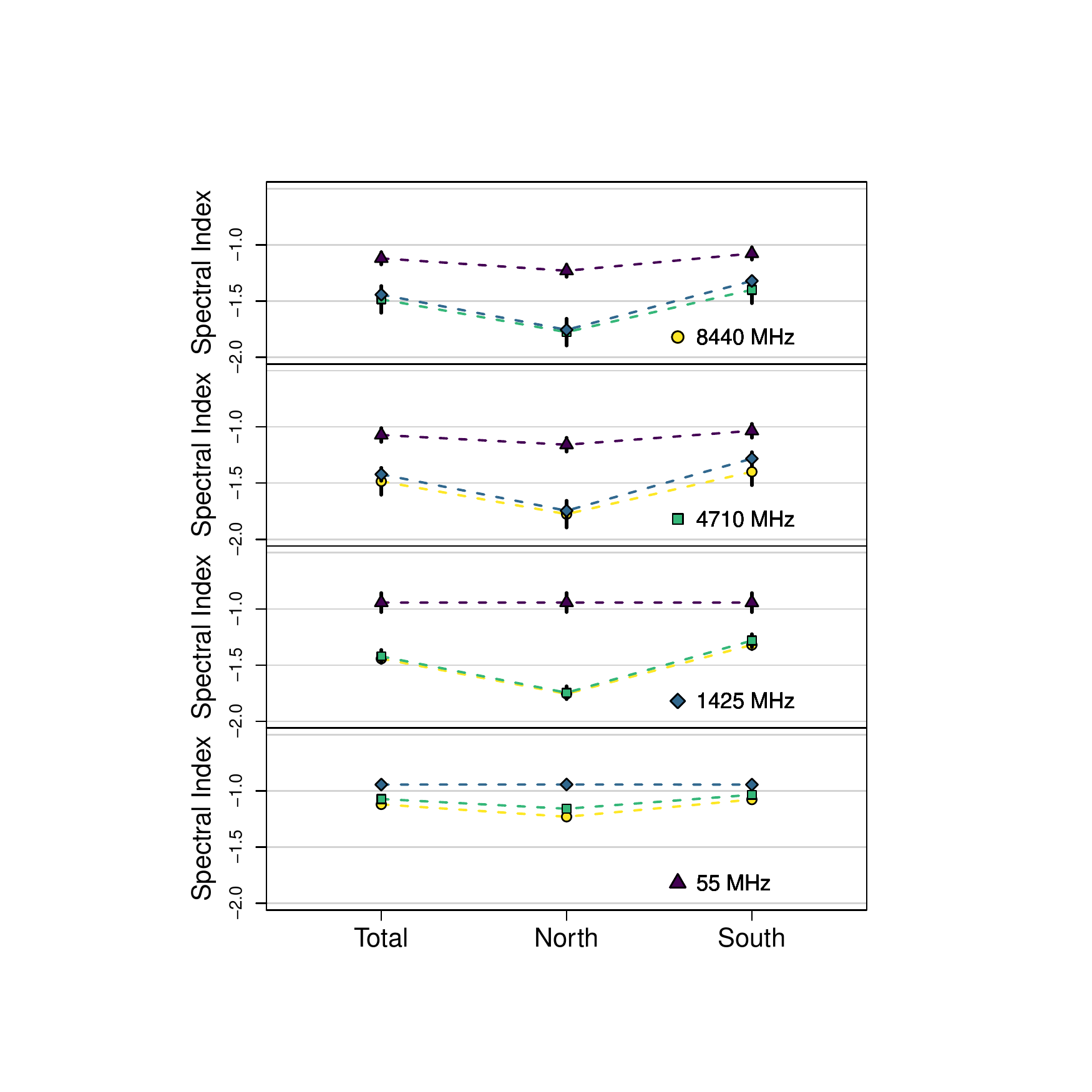}
\caption{\label{fig:si} The point-to-point spectral index values measured from each frequency to all other frequencies in this study. The symbols in all panels of the plot are as follows: 55$\,$MHz -- yellow triangles; 1425$\,$MHz -- green diamonds; 4710$\,$MHz -- blue squares; 8440$\,$MHz -- purple circles. }
\end{figure}

Figure~\ref{fig:si} shows the point-to-point spectral index values measured from each frequency to all  other frequencies in this study. There are several interesting results. 
\begin{enumerate}
\item The spectral index values amongst frequencies $\geq1.4\,$GHz show a steepening high-frequency spectrum. This can be seen most clearly in the second panel from the top of Figure~\ref{fig:si}, where the spectral index from 4.7$\,$GHz to 8.4$\,$GHz is always steeper than the spectral index from 4.7$\,$GHz to 1.4$\,$GHz for all components. 
\item The point-to-point spectral index from 55$\,$MHz to the higher frequencies in this study steepens, i.e. becomes more negative, as the other point increases in frequency. 
This indicates a break frequency between 55$\,$MHz and 1.4$\,$GHz. This indicates either a steepening at high frequencies, a turnover at low frequencies, or a combination of both. However a low-frequency turnover is not observed in the integrated spectrum. Therefore a steepening of the spectra at high frequencies is more likely, which is seen in Figure~\ref{fig:bb}. 
\item The northern lobe always has a spectral index as steep or steeper than the lobe regardless of the frequencies used to measure the spectral index. This suggests a physical difference between the two lobes. 
\end{enumerate}

 These results for the entire spectrum are consistent with a flatter, normal FR$\,$II spectral index coupled with synchrotron losses that steepen the spectra at high frequencies and cause a break frequency at intermediate frequencies (Harwood et al., 2016). The spectral index between 55$\,$MHz and 1.4$\,$GHz is $\alpha=-0.95$ for both lobes. We fit power laws to the lobe spectra for frequencies $>1\,$GHz and found spectral indices of -1.75$\pm0.01$ (northern lobe) and -1.31$\pm0.03$ (southern lobe). Figure~\ref{fig:bb} shows six measurements of the total integrated spectrum at frequencies less than 500$\,$MHz. The spectral index measured from fitting a power law to these points is $\alpha=-0.83$, which we would expect the lobes to mimic if we had more spatially resolved low-frequency measurements.

\begin{table*}
\caption{\label{tab:sp} Source parameters. Uncertainties in the LOFAR measurement are assumed to be 20 per cent. The optical position was converted to J2000 from the B1950 coordinates in \citet{mccarthy91}: B1950 07:31:49.37 +43:50:59.}
\begin{tabular}{lccccc}
 & \multicolumn{2}{c}{\textsc{northern lobe}} & & \multicolumn{2}{c}{\textsc{southern lobe}} \\ \cline{2-3} \cline{5-6} \\[-6pt]
 & $S_{\nu}$  & Offset from &  & $S_{\nu}$  & Offset from \\ 
  & $[$Jy$]$ & host galaxy &  &  $[$Jy$]$ & host galaxy \\ \cline{2-2} \cline{3-3} \cline{5-5} \cline{6-6} \\
 55$\,$MHz &  5.40$\pm$1.1 & 4.34\sarc\  & & 9.53$\pm$1.9  & 4.35\sarc\ \\
1.4$\,$GHz &  0.25$\pm$0.013 & 4.87\sarc\  & & 0.44$\pm$0.022  & 5.03\sarc\ \\
4.7$\,$GHz & 0.031$\pm$1.6$\times$10$^{-3}$ & 4.52\sarc\  & & 0.095$\pm$4.8$\times$10$^{-3}$ & 4.89\sarc\ \\
8.4$\,$GHz & 0.011$\pm$5.5$\times10^{-4}$ & 5.02\sarc\  & & 0.042$\pm$2.1$\times10^{-3}$ & 4.83\sarc\ \\
\end{tabular}
\end{table*}

\section{Discussion}
\label{sec:diss}
The main result is that both the general morphology and spectral index properties of 4C 43.15 are similar to FR$\,$II sources at low redshift. We have determined that 4C 43.15 has historically fallen on the spectral index -- redshift relation because of the steepening of its spectrum at high frequencies, and a break frequency between 55$\,$MHz and 1.4$\,$GHz. The total integrated spectrum has a spectral index of $\alpha=-0.83\pm0.02$ for frequencies below 500$\,$MHz, which is not abnormally steep when compared to other FR$\,$II sources. For example, the median spectral index for the 3CRR sample is $\alpha=-0.8$ \citep{3CRR}. The lowest rest frequency probed is 180$\,$MHz, which is still above where low-frequency turnovers are seen in the spectra of local FR$\,$II sources \citep[e.g.,][]{mckean16,carilli91}. Thus we expect the break frequency to be due to synchrotron losses at high frequencies rather than a low frequency turnover.  

We find no evidence that environmental effects cause a steeper overall spectrum. In fact, the northern lobe, which has the steeper spectral index, is likely undergoing adiabatic expansion into a region of lower density. This is contrary to the scenario discussed by \citet{ak98} where higher ambient densities and temperatures will cause a steeper spectral index. The interaction of 4C 43.15 with its environment will be discussed in detail later in this section. 

The observational bias resulting in the initial classification of 4C 43.15 as having an ultra steep spectrum could be a manifestation of different spectral energy losses at high frequencies when compared to local radio galaxies. It is possible that inverse Compton losses, which scale as (1+$z$)$^4$, combined with spectral ageing, have lowered the break frequency relative to losses from spectral ageing alone. For any two fixed observing frequencies that straddle the break frequency, a lower break frequency will cause a reduction in the intensity measured at the higher frequency, resulting in a steeper measured spectral index. To model the lobe spectra including the contribution from losses due to the CMB, we require spatially resolved measurements at another low frequency (less than $\sim500\,$MHz) to unambiguously determine the low-frequency spectral indices of the lobes of 4C 43.15. We plan to use HBA observations of 4C 43.15 to provide measurements at 150$\,$MHz in future studies. 

In the following subsections we first calculate the apparent ages of the radio lobes and then look at evidence for environmental interaction.  

\subsection{Ages of the radio lobes}
\label{sec:int}
The spectral age can be related to the break frequency $\nu_{br}$ and the magnetic field strength $B$ by:
\begin{equation}
\tau_{\textrm{rad}} = 50.3 \frac{B^{1/2}}{B^2 + B_{\textrm{iC}}^2} [\nu_{br}(1+z)]^{-1/2} \textrm{  Myr}
\end{equation}
\citep[e.g.,][and references therein]{harwood13}. The inverse Compton microwave background radiation has a magnetic field strength $B_{\textrm{iC}}=0.318(1+z)^2$. The units of $B$ and $B_{\textrm{iC}}$ are nT and $\nu_{br}$ is in GHz. Using the standard minimum energy assumptions \cite{carilli97} derived minimum pressures for the hotspots, which correspond to a magnetic field of  32$\,$nT for 4C 43.15, which is consistent with values for Cygnus A \citep{carilli91}. We therefore assume an average value of $B=1\,$nT for the lobes of 4C 43.15, which is consistent with Cygnus A. To calculate $\tau_{\textrm{rad}}$, the break frequency must also be known, and we estimate this from fitting two power laws to integrated flux density measurements: one power low fitted to data at frequencies below 500$\,$MHz, and one power law fitted to data at frequencies above 1$\,$GHz. The frequency at which these two power laws cross is the break frequency. 

Using the spectral indices calculated in the previous section, we estimate the break frequencies of the lobes by finding where the low and high frequency fitted power laws cross. The estimated break frequencies for the northern and southern lobes are 947$\pm12\,$MHz and 662$\pm29\,$MHz, giving apparent ages of 12.7$\pm0.2$ and 15.2$\pm0.7$ Myr, respectively. These ages are reasonable for FR$\,$II sources of this size \citep[e.g.,][]{harwood15}. 

\subsection{Environmental interaction}
\label{sec:env}
The fact that the observed lobes are not the same is clear: the northern lobe is little more than half as bright as the southern lobe, and has a steeper spectral index above 1.4$\,$GHz by $\Delta\alpha=-0.5$. We have thus far found that 4C 43.15 is consistent with local FR$\,$II sources, and therefore we do not expect an internal difference in physical processes driving the two lobes. This suggests there must be an external cause. \citet{humphrey07} found the differences between the lobes in 4C 43.15 to be consistent with orientation effects by modelling Doppler boosting of the hotspots to predict the resulting asymmetry between the lobes for a range of viewing angles and velocities. Although only hot spot advance speeds of 0.4$c$ and viewing angles of $>20\,$deg approach the measured $\Delta\alpha=-0.5$. Since 4C 43.15 is similar to Cygnus A, hot spot advance speeds of $\sim0.05c$ are much more likely. In this scenario, the models in  \citet{humphrey07} predict a value for $\Delta\alpha$ at least an order of magnitude smaller than $-0.5$ for all viewing angles considered. We therefore find it unlikely that orientation is the only cause for the differences between the lobes.

Environmental factors could also cause differences between the lobes. In lower density environments, adiabatic expansion of a radio lobe would lower the surface brightness, effectively shift the break frequency to lower frequencies, and cause a slight steepening of the radio spectrum at higher frequencies. This is consistent with the morphology and spectral index properties of 4C 43.15. The northern lobe is dimmer, appears more diffuse, and has a spectral index steeper than that of the southern lobe. Having ruled out that orientation can explain these asymmetries, this implies that the northern jet is propagating through a lower density medium.

\begin{figure*}
\includegraphics[width=\textwidth,clip,trim=5.5cm 3.5cm 5.5cm 5cm]{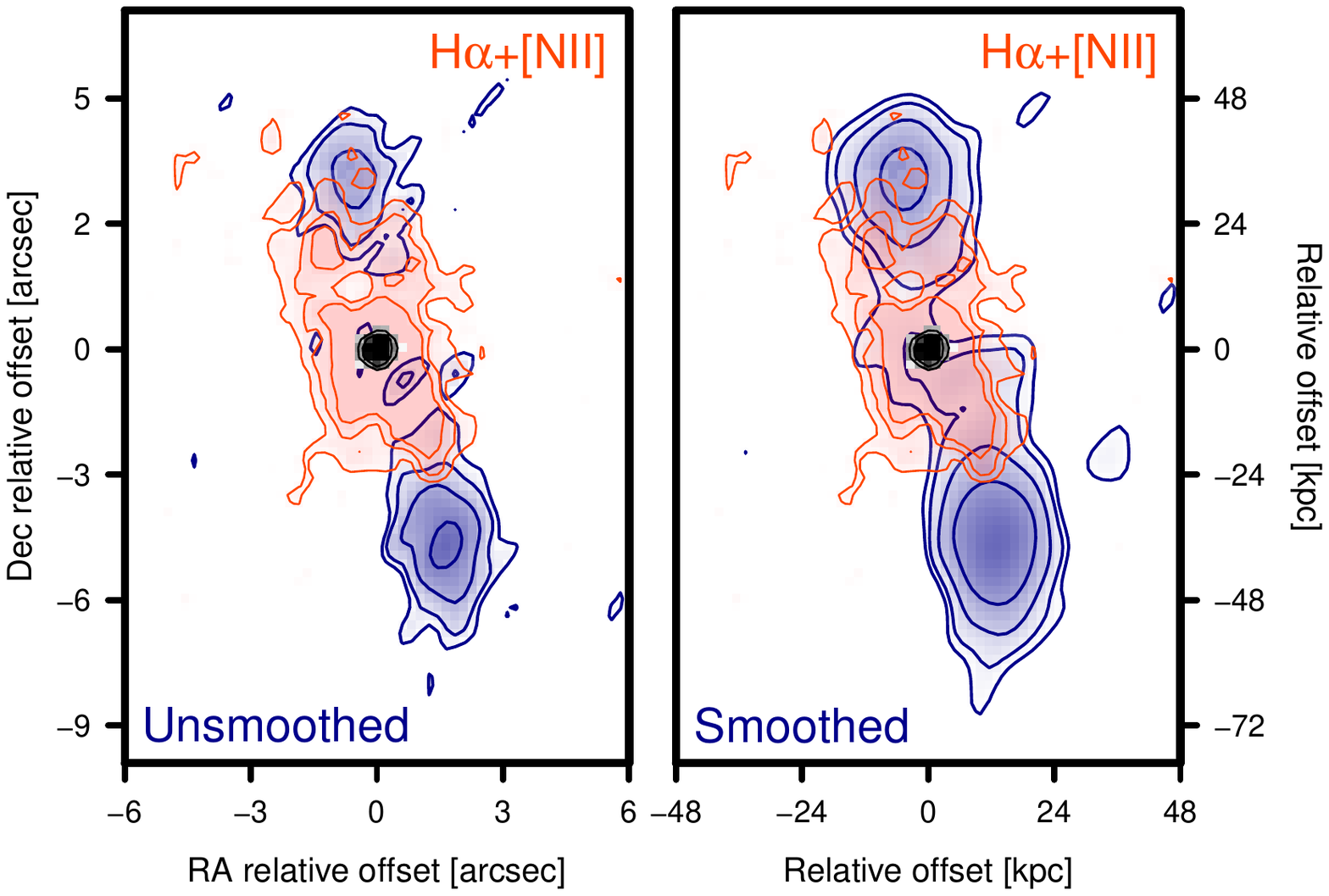}
\caption{\label{fig:env} The spatial distribution of the radio emission compared with the with $K'$-band (2.13$\,\mu$m) continuum from the host galaxy and H$\alpha$+$[$N$\,$II$]$ line emission showing cones of ionized gas \citep{motohara00}. The two panels show the radio images with the same contours as the images in Figure~\ref{fig:tgt} (unsmoothed in the left panel, smoothed in the right panel), overlaid with $K'$-band continuum in black and H$\alpha$+$[$N$\,$II$]$ in red. A separate bright source to the NW has been blanked out. }
\end{figure*}

There is supporting evidence for a lower density medium to the North of the host galaxy. Both Lyman-$\alpha$ \citep{vm03} and H$\alpha$+$[$N$\,$II$]$ \citep{motohara00} are seen to be more extended to the North. Figure~\ref{fig:env} shows the H$\alpha$+$[$N$\,$II$]$ overlaid on the radio images for comparison. Qualitatively the emission line gas is more extended and disturbed towards the North, and reaches farther into the area of the radio lobe. \citet{motohara00} concluded that the Lyman-$\alpha$ and H$\alpha$ emission are both nebular emission from gas ionized by strong UV radiation from the central active galactic nucleus. They estimate the electron density of the ionised gas to be 38$\,$cm$^{-3}$ and 68$\,$cm$^{-3}$ for the northern and southern regions, respectively. The lower density in the North is consistent with adiabatic expansion having a larger impact on the northern lobe relative to the southern lobe. Naively, the ratios between the integrated flux densities of the lobes and the densities of the environment are similar. However determining the expected relationship between the two ratios requires estimating the synchrotron losses from adiabatic expansion, which requires knowing the relevant volumes and densities, then modelling and fully evolving the spectra. Measuring the volumes requires knowing the full extent of the radio emission, which is hard to do if the lobe already has low surface brightness due to adiabatic expansion. This complex modelling is beyond the scope of this paper and will be addressed in future studies (J. Harwood, private communication). 

\section{Conclusions and Outlook}
\label{sec:concl}
We have shown that I-LOFAR LBA is suitable for spatially resolved studies of bright objects. We have presented the first sub-arcsecond image made at frequencies lower than 100$\,$MHz, setting the record for highest spatial resolution at low radio frequencies. This is an exciting prospect that many other science cases will benefit from in the future. 

There are two main conclusions from this study of the spatially resolved low frequency properties of high redshift radio galaxy 4C 43.15: \\

$\bullet $ Low-surface brightness radio emission at low frequencies is seen, for the first time in a high redshift radio galaxy, to be extended between the two radio lobes. The low-frequency morphology is similar to local FR$\,$II radio sources like Cygnus A. 

$\bullet $ The overall spectra for the lobes are ultra steep only when measuring from 55$\,$MHz to frequencies \emph{above} 1.4$\,$GHz. This is likely due to an ultra-steep spectrum at frequencies $\geq1.4\,$GHz with a break frequency between 55$\,$MHz and 1.4$\,$GHz. The low-frequency spectra are consistent with what is found for local FR$\,$II sources. \\
 
This study has revealed that although 4C 43.15 would have been classified as an ultra-steep spectrum source by \citet{db00}, this is likely due to a break frequency at intermediate frequencies, and the spectral index at frequencies less than this break is not abnormally steep for nearby FR$\,$II sources. 
 Steepening of the spectra at high frequencies could be due to synchrotron ageing and inverse Compton losses from the increased magnetic field strength of the cosmic microwave background radiation at higher redshifts.
 Unlike nearby sources, we do not observe curvature in the low frequency spectra, which could be due to the fact that we only observe down to a rest frequency of about 180$\,$MHz. Future observations at 30$\,$MHz (103$\,$MHz rest frequency) or lower would be useful.  

Larger samples with more data points at low to intermediate frequencies are necessary to determine if the observed ultra steep spectra of high redshift radio galaxies also exhibit the same spectral properties as 4C 43.15. 
We will use the methods developed for this paper to study another 10 resolved sources with $2<z<4$, incorporating both LBA and HBA measurements to provide excellent constraints on the low-frequency spectra. 
While a sample size of 11 may not be large enough for general conclusions, it will provide important information on trends in these high redshift sources. These trends can help guide future, large scale studies.

\section*{Acknowledgements}
LKM acknowledges financial support from NWO Top LOFAR project, project no. 614.001.006. 
LKM and HR acknowledge support from the ERC Advanced Investigator programme NewClusters 321271.
The authors would like to thank J. Harwood and H. Intema for many useful discussions. 
RM gratefully acknowledge support from the European Research Council under the European Union's Seventh Framework Programme (FP/2007-2013) /ERC Advanced Grant RADIOLIFE-320745.
This paper is based (in part) on data obtained with the International LOFAR Telescope (ILT). LOFAR (van Haarlem et al. 2013) is the Low Frequency Array designed and constructed by ASTRON. It has facilities in several countries, that are owned by various parties (each with their own funding sources), and that are collectively operated by the ILT foundation under a joint scientific policy.
This research has made use of the NASA/IPAC Extragalactic Database (NED) which is operated by the Jet Propulsion Laboratory, California Institute of Technology, under contract with the National Aeronautics and Space Administration. 
This research made use of Montage. It is funded by the National Science Foundation under Grant Number ACI-1440620, and was previously funded by the National Aeronautics and Space Administration's Earth Science Technology Office, Computation Technologies Project, under Cooperative Agreement Number NCC5-626 between NASA and the California Institute of Technology. 

\bibliographystyle{mnras}
\bibliography{\myreferences}

\begin{thebibliography}{}
\makeatletter
\relax
\def\mn@urlcharsother{\let\do\@makeother \do\$\do\&\do\#\do\^\do\_\do\%\do\~}
\def\mn@doi{\begingroup\mn@urlcharsother \@ifnextchar [ {\mn@doi@}
  {\mn@doi@[]}}
\def\mn@doi@[#1]#2{\def\@tempa{#1}\ifx\@tempa\@empty \href
  {http://dx.doi.org/#2} {doi:#2}\else \href {http://dx.doi.org/#2} {#1}\fi
  \endgroup}
\def\mn@eprint#1#2{\mn@eprint@#1:#2::\@nil}
\def\mn@eprint@arXiv#1{\href {http://arxiv.org/abs/#1} {{\tt arXiv:#1}}}
\def\mn@eprint@dblp#1{\href {http://dblp.uni-trier.de/rec/bibtex/#1.xml}
  {dblp:#1}}
\def\mn@eprint@#1:#2:#3:#4\@nil{\def\@tempa {#1}\def\@tempb {#2}\def\@tempc
  {#3}\ifx \@tempc \@empty \let \@tempc \@tempb \let \@tempb \@tempa \fi \ifx
  \@tempb \@empty \def\@tempb {arXiv}\fi \@ifundefined
  {mn@eprint@\@tempb}{\@tempb:\@tempc}{\expandafter \expandafter \csname
  mn@eprint@\@tempb\endcsname \expandafter{\@tempc}}}

\bibitem[\protect\citeauthoryear{{Athreya} \& {Kapahi}}{{Athreya} \&
  {Kapahi}}{1998}]{ak98}
{Athreya} R.~M.,  {Kapahi} V.~K.,  1998, \mn@doi [Journal of Astrophysics and
  Astronomy] {10.1007/BF02714911}, \href
  {http://adsabs.harvard.edu/abs/1998JApA...19...63A} {19, 63}

\bibitem[\protect\citeauthoryear{{Becker}, {White}  \& {Edwards}}{{Becker}
  et~al.}{1991}]{becker91}
{Becker} R.~H.,  {White} R.~L.,   {Edwards} A.~L.,  1991, \mn@doi [\apjs]
  {10.1086/191529}, \href {http://adsabs.harvard.edu/abs/1991ApJS...75....1B}
  {75, 1}

\bibitem[\protect\citeauthoryear{{Best}, {Longair}  \& {R{\"o}ttgering}}{{Best}
  et~al.}{1997}]{blr97}
{Best} P.~N.,  {Longair} M.~S.,   {R{\"o}ttgering} H.~J.~A.,  1997, ArXiv
  Astrophysics e-prints, \href
  {http://adsabs.harvard.edu/abs/1997astro.ph.11010B} {}

\bibitem[\protect\citeauthoryear{{Blumenthal} \& {Miley}}{{Blumenthal} \&
  {Miley}}{1979}]{bm79}
{Blumenthal} G.,  {Miley} G.,  1979, \aap, \href
  {http://adsabs.harvard.edu/abs/1979A%26A....80...13B} {80, 13}

\bibitem[\protect\citeauthoryear{{Blundell}, {Rawlings}  \&
  {Willott}}{{Blundell} et~al.}{1999}]{blundell99}
{Blundell} K.~M.,  {Rawlings} S.,   {Willott} C.~J.,  1999, \mn@doi [\aj]
  {10.1086/300721}, \href {http://adsabs.harvard.edu/abs/1999AJ....117..677B}
  {117, 677}

\bibitem[\protect\citeauthoryear{{Bridle} \& {Schwab}}{{Bridle} \&
  {Schwab}}{1999}]{bridleschwab}
{Bridle} A.~H.,  {Schwab} F.~R.,  1999, in {Taylor} G.~B.,  {Carilli} C.~L.,
  {Perley} R.~A.,  eds,  Astronomical Society of the Pacific Conference Series
  Vol. 180, Synthesis Imaging in Radio Astronomy II. p.~371

\bibitem[\protect\citeauthoryear{{Carilli}, {Perley}, {Dreher}  \&
  {Leahy}}{{Carilli} et~al.}{1991}]{carilli91}
{Carilli} C.~L.,  {Perley} R.~A.,  {Dreher} J.~W.,   {Leahy} J.~P.,  1991,
  \mn@doi [\apj] {10.1086/170813}, \href
  {http://adsabs.harvard.edu/abs/1991ApJ...383..554C} {383, 554}

\bibitem[\protect\citeauthoryear{{Carilli}, {R{\"o}ttgering}, {van Ojik},
  {Miley}, {Breugel}  \& {W.~J.~M.~van}}{{Carilli} et~al.}{1997}]{carilli97}
{Carilli} C.~L.,  {R{\"o}ttgering} H.~J.~A.,  {van Ojik} R.,  {Miley} G.~K.,
  {Breugel}  {W.~J.~M.~van} 1997, \mn@doi [\apjs] {10.1086/312973}, \href
  {http://adsabs.harvard.edu/abs/1997ApJS..109....1C} {109, 1}

\bibitem[\protect\citeauthoryear{{Chambers}, {Miley}  \& {van
  Breugel}}{{Chambers} et~al.}{1987}]{cmb87}
{Chambers} K.~C.,  {Miley} G.~K.,   {van Breugel} W.,  1987, \mn@doi [\nat]
  {10.1038/329604a0}, \href {http://adsabs.harvard.edu/abs/1987Natur.329..604C}
  {329, 604}

\bibitem[\protect\citeauthoryear{{Chambers}, {Miley}  \& {van
  Breugel}}{{Chambers} et~al.}{1990}]{chambers90}
{Chambers} K.~C.,  {Miley} G.~K.,   {van Breugel} W.~J.~M.,  1990, \mn@doi
  [\apj] {10.1086/169316}, \href
  {http://adsabs.harvard.edu/abs/1990ApJ...363...21C} {363, 21}

\bibitem[\protect\citeauthoryear{{Cohen}, {Lane}, {Cotton}, {Kassim}, {Lazio},
  {Perley}, {Condon}  \& {Erickson}}{{Cohen} et~al.}{2007}]{cohen07}
{Cohen} A.~S.,  {Lane} W.~M.,  {Cotton} W.~D.,  {Kassim} N.~E.,  {Lazio}
  T.~J.~W.,  {Perley} R.~A.,  {Condon} J.~J.,   {Erickson} W.~C.,  2007,
  \mn@doi [\aj] {10.1086/520719}, \href
  {http://adsabs.harvard.edu/abs/2007AJ....134.1245C} {134, 1245}

\bibitem[\protect\citeauthoryear{{Condon}, {Cotton}, {Greisen}, {Yin},
  {Perley}, {Taylor}  \& {Broderick}}{{Condon} et~al.}{1998}]{nvss}
{Condon} J.~J.,  {Cotton} W.~D.,  {Greisen} E.~W.,  {Yin} Q.~F.,  {Perley}
  R.~A.,  {Taylor} G.~B.,   {Broderick} J.~J.,  1998, \mn@doi [\aj]
  {10.1086/300337}, \href {http://adsabs.harvard.edu/abs/1998AJ....115.1693C}
  {115, 1693}

\bibitem[\protect\citeauthoryear{{Cotton}}{{Cotton}}{1995}]{cotton95}
{Cotton} W.~D.,  1995, in {Zensus} J.~A.,  {Diamond} P.~J.,   {Napier} P.~J.,
  eds,  Astronomical Society of the Pacific Conference Series Vol. 82, Very
  Long Baseline Interferometry and the VLBA. p.~189

\bibitem[\protect\citeauthoryear{{De Breuck}, {van Breugel}, {R{\"o}ttgering}
  \& {Miley}}{{De Breuck} et~al.}{2000}]{db00}
{De Breuck} C.,  {van Breugel} W.,  {R{\"o}ttgering} H.~J.~A.,   {Miley} G.,
  2000, \mn@doi [\aaps] {10.1051/aas:2000181}, \href
  {http://adsabs.harvard.edu/abs/2000A%26AS..143..303D} {143, 303}

\bibitem[\protect\citeauthoryear{{Douglas}, {Bash}, {Bozyan}, {Torrence}  \&
  {Wolfe}}{{Douglas} et~al.}{1996}]{texas}
{Douglas} J.~N.,  {Bash} F.~N.,  {Bozyan} F.~A.,  {Torrence} G.~W.,   {Wolfe}
  C.,  1996, \mn@doi [\aj] {10.1086/117932}, \href
  {http://adsabs.harvard.edu/abs/1996AJ....111.1945D} {111, 1945}

\bibitem[\protect\citeauthoryear{{Fanaroff} \& {Riley}}{{Fanaroff} \&
  {Riley}}{1974}]{fr74}
{Fanaroff} B.~L.,  {Riley} J.~M.,  1974, \mn@doi [\mnras]
  {10.1093/mnras/167.1.31P}, \href
  {http://adsabs.harvard.edu/abs/1974MNRAS.167P..31F} {167, 31P}

\bibitem[\protect\citeauthoryear{{Ficarra}, {Grueff}  \&
  {Tomassetti}}{{Ficarra} et~al.}{1985}]{bologna}
{Ficarra} A.,  {Grueff} G.,   {Tomassetti} G.,  1985, \aaps, \href
  {http://adsabs.harvard.edu/abs/1985A%26AS...59..255F} {59, 255}

\bibitem[\protect\citeauthoryear{{Gower}, {Scott}  \& {Wills}}{{Gower}
  et~al.}{1967}]{gower67}
{Gower} J.~F.~R.,  {Scott} P.~F.,   {Wills} D.,  1967, Memoirs of the RAS,
  \href {http://adsabs.harvard.edu/abs/1967MmRAS..71...49G} {71, 49}

\bibitem[\protect\citeauthoryear{{Greisen}}{{Greisen}}{2003}]{greisen03}
{Greisen} E.~W.,  2003, \mn@doi [Information Handling in Astronomy - Historical
  Vistas] {10.1007/0-306-48080-8_7}, \href
  {http://adsabs.harvard.edu/abs/2003ASSL..285..109G} {285, 109}

\bibitem[\protect\citeauthoryear{{Hales}, {Baldwin}  \& {Warner}}{{Hales}
  et~al.}{1993}]{6c}
{Hales} S.~E.~G.,  {Baldwin} J.~E.,   {Warner} P.~J.,  1993, \mn@doi [\mnras]
  {10.1093/mnras/263.1.25}, \href
  {http://adsabs.harvard.edu/abs/1993MNRAS.263...25H} {263, 25}

\bibitem[\protect\citeauthoryear{{Harwood}, {Hardcastle}, {Croston}  \&
  {Goodger}}{{Harwood} et~al.}{2013}]{harwood13}
{Harwood} J.~J.,  {Hardcastle} M.~J.,  {Croston} J.~H.,   {Goodger} J.~L.,
  2013, \mn@doi [\mnras] {10.1093/mnras/stt1526}, \href
  {http://adsabs.harvard.edu/abs/2013MNRAS.435.3353H} {435, 3353}

\bibitem[\protect\citeauthoryear{{Harwood}, {Hardcastle}  \&
  {Croston}}{{Harwood} et~al.}{2015}]{harwood15}
{Harwood} J.~J.,  {Hardcastle} M.~J.,   {Croston} J.~H.,  2015, \mn@doi
  [\mnras] {10.1093/mnras/stv2194}, \href
  {http://adsabs.harvard.edu/abs/2015MNRAS.454.3403H} {454, 3403}

\bibitem[\protect\citeauthoryear{{Humphrey}, {Villar-Mart{\'{\i}}n}, {Fosbury},
  {Binette}, {Vernet}, {De Breuck}  \& {di Serego Alighieri}}{{Humphrey}
  et~al.}{2007}]{humphrey07}
{Humphrey} A.,  {Villar-Mart{\'{\i}}n} M.,  {Fosbury} R.,  {Binette} L.,
  {Vernet} J.,  {De Breuck} C.,   {di Serego Alighieri} S.,  2007, \mn@doi
  [\mnras] {10.1111/j.1365-2966.2006.11344.x}, \href
  {http://adsabs.harvard.edu/abs/2007MNRAS.375..705H} {375, 705}

\bibitem[\protect\citeauthoryear{{Intema}, {Jagannathan}, {Mooley}  \&
  {Frail}}{{Intema} et~al.}{2016}]{tgssadr}
{Intema} H.~T.,  {Jagannathan} P.,  {Mooley} K.~P.,   {Frail} D.~A.,  2016,
  preprint, \href {http://adsabs.harvard.edu/abs/2016arXiv160304368I} {}
  (\mn@eprint {arXiv} {1603.04368})

\bibitem[\protect\citeauthoryear{{Klamer}, {Ekers}, {Bryant}, {Hunstead},
  {Sadler}  \& {De Breuck}}{{Klamer} et~al.}{2006}]{klamer06}
{Klamer} I.~J.,  {Ekers} R.~D.,  {Bryant} J.~J.,  {Hunstead} R.~W.,  {Sadler}
  E.~M.,   {De Breuck} C.,  2006, \mn@doi [\mnras]
  {10.1111/j.1365-2966.2006.10714.x}, \href
  {http://adsabs.harvard.edu/abs/2006MNRAS.371..852K} {371, 852}

\bibitem[\protect\citeauthoryear{{Laing}, {Riley}  \& {Longair}}{{Laing}
  et~al.}{1983}]{3CRR}
{Laing} R.~A.,  {Riley} J.~M.,   {Longair} M.~S.,  1983, \mn@doi [\mnras]
  {10.1093/mnras/204.1.151}, \href
  {http://adsabs.harvard.edu/abs/1983MNRAS.204..151L} {204, 151}

\bibitem[\protect\citeauthoryear{{Linsky}, {Rickett}  \& {Redfield}}{{Linsky}
  et~al.}{2008}]{linsky08}
{Linsky} J.~L.,  {Rickett} B.~J.,   {Redfield} S.,  2008, \mn@doi [\apj]
  {10.1086/526420}, \href {http://adsabs.harvard.edu/abs/2008ApJ...675..413L}
  {675, 413}

\bibitem[\protect\citeauthoryear{{McCarthy}}{{McCarthy}}{1991}]{mccarthy91}
{McCarthy} P.~J.,  1991, \mn@doi [\aj] {10.1086/115890}, \href
  {http://adsabs.harvard.edu/abs/1991AJ....102..518M} {102, 518}

\bibitem[\protect\citeauthoryear{{McKean et al.}}{{McKean et
  al.}}{2016}]{mckean16}
{McKean et al.} 2016, {\mnras}, submitted

\bibitem[\protect\citeauthoryear{{McMullin}, {Waters}, {Schiebel}, {Young}  \&
  {Golap}}{{McMullin} et~al.}{2007}]{casa}
{McMullin} J.~P.,  {Waters} B.,  {Schiebel} D.,  {Young} W.,   {Golap} K.,
  2007, in {Shaw} R.~A.,  {Hill} F.,   {Bell} D.~J.,  eds,  Astronomical
  Society of the Pacific Conference Series Vol. 376, Astronomical Data Analysis
  Software and Systems XVI. p.~127

\bibitem[\protect\citeauthoryear{{Miley}}{{Miley}}{1968}]{miley68}
{Miley} G.~K.,  1968, \mn@doi [\nat] {10.1038/218933a0}, \href
  {http://adsabs.harvard.edu/abs/1968Natur.218..933M} {218, 933}

\bibitem[\protect\citeauthoryear{{Miley} \& {De Breuck}}{{Miley} \& {De
  Breuck}}{2008}]{mdb08}
{Miley} G.,  {De Breuck} C.,  2008, \mn@doi [\aapr]
  {10.1007/s00159-007-0008-z}, \href
  {http://adsabs.harvard.edu/abs/2008A%26ARv..15...67M} {15, 67}

\bibitem[\protect\citeauthoryear{{Mold{\'o}n} et~al.,}{{Mold{\'o}n}
  et~al.}{2015}]{moldon15}
{Mold{\'o}n} J.,  et~al., 2015, \mn@doi [\aap] {10.1051/0004-6361/201425042},
  \href {http://adsabs.harvard.edu/abs/2015A%26A...574A..73M} {574, A73}

\bibitem[\protect\citeauthoryear{{Moran} \& {Dhawan}}{{Moran} \&
  {Dhawan}}{1995}]{moran95}
{Moran} J.~M.,  {Dhawan} V.,  1995, in {Zensus} J.~A.,  {Diamond} P.~J.,
  {Napier} P.~J.,  eds,  Astronomical Society of the Pacific Conference Series
  Vol. 82, Very Long Baseline Interferometry and the VLBA. p.~161

\bibitem[\protect\citeauthoryear{{Motohara} et~al.,}{{Motohara}
  et~al.}{2000}]{motohara00}
{Motohara} K.,  et~al., 2000, \mn@doi [\pasj] {10.1093/pasj/52.1.33}, \href
  {http://adsabs.harvard.edu/abs/2000PASJ...52...33M} {52, 33}

\bibitem[\protect\citeauthoryear{{Neeser}, {Eales}, {Law-Green}, {Leahy}  \&
  {Rawlings}}{{Neeser} et~al.}{1995}]{neeser95}
{Neeser} M.~J.,  {Eales} S.~A.,  {Law-Green} J.~D.,  {Leahy} J.~P.,
  {Rawlings} S.,  1995, \mn@doi [\apj] {10.1086/176201}, \href
  {http://adsabs.harvard.edu/abs/1995ApJ...451...76N} {451, 76}

\bibitem[\protect\citeauthoryear{{Offringa}}{{Offringa}}{2010}]{offringa10}
{Offringa} A.~R.,  2010, {AOFlagger: RFI Software}, Astrophysics Source Code
  Library (\mn@eprint {ascl} {1010.017})

\bibitem[\protect\citeauthoryear{{Pauliny-Toth}, {Wade}  \&
  {Heeschen}}{{Pauliny-Toth} et~al.}{1966}]{pt66}
{Pauliny-Toth} I.~I.~K.,  {Wade} C.~M.,   {Heeschen} D.~S.,  1966, \mn@doi
  [\apjs] {10.1086/190137}, \href
  {http://adsabs.harvard.edu/abs/1966ApJS...13...65P} {13, 65}

\bibitem[\protect\citeauthoryear{{Pentericci}, {Van Reeven}, {Carilli},
  {R{\"o}ttgering}  \& {Miley}}{{Pentericci} et~al.}{2000a}]{pentVLA00}
{Pentericci} L.,  {Van Reeven} W.,  {Carilli} C.~L.,  {R{\"o}ttgering}
  H.~J.~A.,   {Miley} G.~K.,  2000a, \mn@doi [\aaps] {10.1051/aas:2000104},
  \href {http://adsabs.harvard.edu/abs/2000A%26AS..145..121P} {145, 121}

\bibitem[\protect\citeauthoryear{{Pentericci} et~al.,}{{Pentericci}
  et~al.}{2000b}]{pent00}
{Pentericci} L.,  et~al., 2000b, \aap, \href
  {http://adsabs.harvard.edu/abs/2000A%26A...361L..25P} {361, L25}

\bibitem[\protect\citeauthoryear{{Planck Collaboration} et~al.,}{{Planck
  Collaboration} et~al.}{2015}]{planckcosmo}
{Planck Collaboration} et~al., 2015, preprint, \href
  {http://adsabs.harvard.edu/abs/2015arXiv150201589P} {} (\mn@eprint {arXiv}
  {1502.01589})

\bibitem[\protect\citeauthoryear{{Quirrenbach}}{{Quirrenbach}}{1992}]{quirrenbach92}
{Quirrenbach} A.,  1992, in {Klare} G.,  ed.,  Reviews in Modern Astronomy Vol.
  5, {Variability and VLBI Observations of Extragalactic Radio Surces}. pp
  214--228

\bibitem[\protect\citeauthoryear{{Rickett}}{{Rickett}}{1986}]{rickett86}
{Rickett} B.~J.,  1986, \mn@doi [\apj] {10.1086/164444}, \href
  {http://adsabs.harvard.edu/abs/1986ApJ...307..564R} {307, 564}

\bibitem[\protect\citeauthoryear{{R\"{o}ttgering}, {Lacy}, {Miley}, {Chambers}
  \& {Saunders}}{{R\"{o}ttgering} et~al.}{1994}]{rott94}
{R\"{o}ttgering} H.~J.~A.,  {Lacy} M.,  {Miley} G.~K.,  {Chambers} K.~C.,
  {Saunders} R.,  1994, \aaps, \href
  {http://adsabs.harvard.edu/abs/1994A%26AS..108...79R} {108}

\bibitem[\protect\citeauthoryear{{Salvini} \& {Wijnholds}}{{Salvini} \&
  {Wijnholds}}{2014}]{sw14}
{Salvini} S.,  {Wijnholds} S.~J.,  2014, \mn@doi [\aap]
  {10.1051/0004-6361/201424487}, \href
  {http://adsabs.harvard.edu/abs/2014A%26A...571A..97S} {571, A97}

\bibitem[\protect\citeauthoryear{{Scaife} \& {Heald}}{{Scaife} \&
  {Heald}}{2012}]{sh12}
{Scaife} A.~M.~M.,  {Heald} G.~H.,  2012, \mn@doi [\mnras]
  {10.1111/j.1745-3933.2012.01251.x}, \href
  {http://adsabs.harvard.edu/abs/2012MNRAS.423L..30S} {423, L30}

\bibitem[\protect\citeauthoryear{{Thompson}, {Moran}  \& {Swenson}}{{Thompson}
  et~al.}{2001}]{tms01}
{Thompson} A.~R.,  {Moran} J.~M.,   {Swenson} Jr. G.~W.,  2001, {Interferometry
  and Synthesis in Radio Astronomy, 2nd Edition}.
2nd ed.~ New York : Wiley, c2001.xxiii, 692 p.~: ill.~; 25 cm.~''A
  Wiley-Interscience publication.'' Includes bibliographical references and
  indexes.~ISBN : 0471254924''

\bibitem[\protect\citeauthoryear{{Tielens}, {Miley}  \& {Willis}}{{Tielens}
  et~al.}{1979}]{tielens79}
{Tielens} A.~G.~G.~M.,  {Miley} G.~K.,   {Willis} A.~G.,  1979, \aaps, \href
  {http://adsabs.harvard.edu/abs/1979A%26AS...35..153T} {35, 153}

\bibitem[\protect\citeauthoryear{{Varenius} et~al.,}{{Varenius}
  et~al.}{2015}]{varenius15}
{Varenius} E.,  et~al., 2015, \mn@doi [\aap] {10.1051/0004-6361/201425089},
  \href {http://adsabs.harvard.edu/abs/2015A%26A...574A.114V} {574, A114}

\bibitem[\protect\citeauthoryear{{Villar-Mart{\'{\i}}n}, {Vernet}, {di Serego
  Alighieri}, {Fosbury}, {Humphrey}  \& {Pentericci}}{{Villar-Mart{\'{\i}}n}
  et~al.}{2003}]{vm03}
{Villar-Mart{\'{\i}}n} M.,  {Vernet} J.,  {di Serego Alighieri} S.,  {Fosbury}
  R.,  {Humphrey} A.,   {Pentericci} L.,  2003, \mn@doi [\mnras]
  {10.1046/j.1365-2966.2003.07090.x}, \href
  {http://adsabs.harvard.edu/abs/2003MNRAS.346..273V} {346, 273}

\bibitem[\protect\citeauthoryear{{Wardle} \& {Miley}}{{Wardle} \&
  {Miley}}{1974}]{wm74}
{Wardle} J.~F.~C.,  {Miley} G.~K.,  1974, \aap, \href
  {http://adsabs.harvard.edu/abs/1974A%26A....30..305W} {30, 305}

\bibitem[\protect\citeauthoryear{{White} \& {Becker}}{{White} \&
  {Becker}}{1992}]{wb92}
{White} R.~L.,  {Becker} R.~H.,  1992, \mn@doi [\apjs] {10.1086/191656}, \href
  {http://adsabs.harvard.edu/abs/1992ApJS...79..331W} {79, 331}

\bibitem[\protect\citeauthoryear{{Wucknitz}}{{Wucknitz}}{2010}]{wucknitz10}
{Wucknitz} O.,  2010, in ISKAF2010 Science Meeting. p.~58 (\mn@eprint {arXiv}
  {1008.4358})

\bibitem[\protect\citeauthoryear{{van Haarlem} et~al.,}{{van Haarlem}
  et~al.}{2013}]{vh13}
{van Haarlem} M.~P.,  et~al., 2013, \mn@doi [\aap]
  {10.1051/0004-6361/201220873}, \href
  {http://adsabs.harvard.edu/abs/2013A%26A...556A...2V} {556, A2}

\bibitem[\protect\citeauthoryear{{van der Tol}, {Jeffs}  \& {van der
  Veen}}{{van der Tol} et~al.}{2007}]{vdt07}
{van der Tol} S.~.,  {Jeffs} B.~D.,   {van der Veen} A.-J.~.,  2007, \mn@doi
  [IEEE Transactions on Signal Processing] {10.1109/TSP.2007.896243}, \href
  {http://adsabs.harvard.edu/abs/2007ITSP...55.4497V} {55, 4497}

\makeatother
\end{thebibliography}

\label{lastpage}

\end{document}